\documentclass[]{aiaa-tc}

\AIAApapernumber{YEAR-NUMBER}
\AIAAconference{Conference Name, Date, and Location}
\AIAAcopyright{\AIAAcopyrightD{YEAR}}

\usepackage{graphicx}
\usepackage{epsfig}
\usepackage{color}
\usepackage{psfrag}
\usepackage{subfigure}
\usepackage{subfigmat}
\usepackage{subeqnarray}
\usepackage{hyperref}
\usepackage{amsmath}
\usepackage{amsfonts,stmaryrd,amssymb,theorem}
\usepackage{xcolor,colortbl}
\usepackage[mathscr]{euscript}
\usepackage[ruled]{algorithm2e}
\usepackage{algorithmic}
\usepackage{changepage}
\usepackage{multirow}
\graphicspath{{../Figures/}{./Figure/}}
\usepackage{rotating}
\usepackage{bm}
\usepackage{lineno}
\usepackage{textgreek}
\usepackage{enumitem}
\usepackage{array}
\usepackage{tikz}

\usepackage[capitalize]{cleveref}

\usepackage[export]{adjustbox}

\graphicspath{{./figures/}}

\title{Numerical framework for transcritical real-fluid reacting flow simulations using the flamelet progress variable approach}

\author{
    Peter C. Ma\thanks{Graduate Research Assistant, Department of Mechanical Engineering.} \ and Daniel T. Banuti\thanks{Postdoctoral Research Fellow, Center for Turbulence Research.}\\
    {\normalsize\itshape Stanford University, Stanford, CA 94305, USA}\\
    \and
    Jean-Pierre Hickey\thanks{Assistant Professor, Department of Mechanical and Mechatronics Engineering.}\\
    {\normalsize\itshape University of Waterloo, Waterloo, ON N2L 3G1, Canada}\\
    \and
    Matthias Ihme\thanks{Assistant Professor, Department of Mechanical Engineering.}\\
    {\normalsize\itshape Stanford University, Stanford, CA 94305, USA}
}

\begin{document}

\maketitle

\begin{abstract}
An extension to the classical FPV model is developed for transcritical real-fluid combustion simulations in the context of finite volume, fully compressible, explicit solvers. A double-flux model is developed for transcritical flows to eliminate the spurious pressure oscillations. A hybrid scheme with entropy-stable flux correction is formulated to robustly represent large density ratios. The thermodynamics for ideal-gas values is modeled by a linearized specific heat ratio model. Parameters needed for the cubic EoS are pre-tabulated for the evaluation of departure functions and a quadratic expression is used to recover the attraction parameter. The novelty of the proposed approach lies in the ability to account for pressure and temperature variations from the baseline table. Cryogenic LOX/GH2 mixing and reacting cases are performed to demonstrate the capability of the proposed approach in multidimensional simulations. The proposed combustion model and numerical schemes are directly applicable for LES simulations of real applications under transcritical conditions.
\end{abstract}

\section{Introduction}
Liquid rocket engines (LRE) are one of the many practical applications which operate near or above the critical point of the working fluid.  The large expansion ratio needed to generate the required thrust means that the combustion chamber operates at extremely high pressures, typically between 30 and 200 bar. In liquid rocket engines, the high mass flow rate is achieved by injecting high energy density cryogenic fuel and oxidizer into the combustion chamber through an extensive array of coaxial or impinging jets, see Cheroudi~\cite{Chehroudi2012} for a review of high-pressure injection strategies. In the most common coaxial setup, the oxidizer generally exits the lip of the injectors below the critical temperature while being above the critical pressure of the fluid. In order to initiate chemical reactions for combustion, the cryogenic fluid must undergo a transcritical phase change to a supercritical state. In the transcritical regime, the thermo-physical properties of the fluid undergoes drastic changes for minute perturbations of the baseline thermodynamic state. This highly non-linear flow behavior requires the use of a generalized state equation to account for the complex thermodynamic properties.

A physics-based understanding of real fluid effects is needed to fully address the modeling challenges for trans- and supercritical combustion flow. Early experiments sought to address the physics of a sub- to supercritical jet phase change. The transcritical effects are most clearly observed in the visualization of transcritical pure mixing   \cite{Mayer1998,Oschwald1999,Chehroudi2002,Habiballah2006}. At subcritical pressures, a liquid jet undergoes a convection driven process of atomization and breakup. At supercritical pressure, the same liquid jet undergoes a diffusion driven mixing \textemdash primarily due to the lack of surface tension, increased diffusivity and reduction in evaporation enthalpy.  In the diffuse region between the liquid core and the gaseous outer flow, large thermo-physical gradients are observed and the dense core length is significantly reduced under supercritical pressures \cite{Davis2007}.  Under these conditions, a dense gas/gas mixing of the propellants and oxidizers is typically observed. The delineation between atomization and diffusion  driven mixing of multi-species mixtures has been characterized by Dahms and Oefelein \cite{Dahms2013}. 
     
  
In high pressure rocket combustion chambers, the turbulence time scale has a magnitude of 1$\mu s$ in the reactive shear layer while the length scale is 1$\mu m$ near the exit lip~\cite{Ivancic2002}. The flame thickness, for non-premixed combustion,  decreases with the inverse of the square root of pressure and strain rate~\cite{Ribert2008}. In a follow-up study, Lacaze~and~Oefelein~\cite{Lacaze2012} suggested that the flame thickness is proportional to the inverse of the square root of pressure. The basis of the flamelet formulation~\cite{Pitsch2006} rests on a scale separation between the turbulence and the chemical scales in both time and space. Ivancic~and~Mayer~\cite{Ivancic2002} inferred that the turbulent and chemical time scales could be of similar magnitude \textemdash a result that could invalidate the laminar flamelet assumption. To address this issue, Zong~et~al.~\cite{Zong2008} showed the appropriateness of the flamelet assumptions in coaxial injectors in the supercritical regime using scaling arguments. The validity of the flamelet approach has lead to a number of flamelet-based numerically studied~\cite{Cutrone2010, Ribert2008, Kim2011a, Lacaze2012, huo2014general, banuti2016sub}.

Pre-tabulated flamelet approaches  have been successfully applied to a variety of combustion cases~\cite{Pitsch1998,Pitsch2006}. The original look-up tables based on mixture fraction, $\widetilde{Z}$, and its variance, $\widetilde{Z^{''2}}$, tend to be inadequate for describing topologically complex flames. This is particularly true for lifted flames where the mixture fraction alone is insufficient to model the full physical complexity~\cite{wu2015pareto,wu2016compliance}.  Careful experiments in high-pressure combustors have revealed a multiplicity of flame anchoring and stabilization mechanisms~\cite{Candel2006,Yang2007}, sometimes with a lifted flame configuration. In order to capture the intrinsic physics of these flames using a computationally tractable combustion modeling approach, a progress variable needs to be transported in addition to the mixture fraction. The flamelet progress variable (FPV) approach~\cite{Pierce2004,Ihme2005} has been shown to better capture the complex physics in detached flames. The FPV approach has been used for supercritical simulations by Cutrone~et~al.~\cite{Cutrone2010}  and more recently by Giorgi~et~al.~\cite{Giorgi2014} in the context of Reynolds averaged simulations. The extension of the FPV model for trans- and supercritical flows in the fully compressible context with large-eddy simulation (LES) remains, for the most part, unreported, specifically with regards to the pressure and temperature coupling.
  
The use of LES for trans- and supercritical flows has been initially explored using one-step chemistry~\cite{Oefelein1998} and more recently extended for hydrogen combustion by Schmitt~et~al.~\cite{Schmitt2011} using a thickened flame approach. Other groups have proposed an extension to the linear eddy model to account for combustion in high-pressure systems~\cite{masquelet2010large}.  Further considerations have been proposed to account for subgrid-scale modeling under supercritical conditions~\cite{Selle2007,Huo2013b}, sensitivity of the state equation \cite{Petit2013b}, numerical stability issues~\cite{Terashima2011,Terashima2012a,Terashima2013, Lacaze2013,ma2014supercritical,ma2016numerical,ma2016entropy}, non-reflecting boundary conditions~\cite{Okongo2002,Coussement2013}, and compressibility effects~\cite{petit2015framework}. While direct numerical simulations (DNS) have been successfully applied to trans- and supercritical flows in idealized configurations~\cite{Miller2001,Okongo2002a,Okongo2003a,Taskinoglu2010}, the computational limitations restrict the applications to academic problems.
      
In this work, we propose an extension to the classical FPV approach~\cite{Pierce2004, Ihme2005} for trans- and supercritical combustion simulations in the context of finite volume, fully compressible, explicit solvers. The novelty of the present work lies in the ability to account for pressure and temperature variations from the baseline tabulated values using a computationally tractable pre-tabulated combustion chemistry in a thermodynamically consistent fashion. In addition, we show that the solution of the laminar flamelets in mixture fraction space and the chemistry tabulation requires special considerations in order to fully model the non-linear effects in transcritical flows.

\section{Thermodynamics\label{sec:RealFluid}}

\subsection{Equation of state}
Cubic equations of state offer an acceptable compromise between the conflicting requirements of accuracy and computational tractability.~\cite{miller2001direct}. The Peng-Robinson (PR) cubic EoS~\cite{peng1976new} is used in this study for the evaluation of thermodynamic quantities, which can be written as
\begin{equation}
    \label{EQ_PR_EQUATION_P}
    p = \frac{R T}{v - b} - \frac{a}{v^2+2bv-b^2}
\end{equation}
where $p$ is the pressure, $R$ is the gas constant, $T$ is the temperature, $v$ is the specific volume, and the attraction parameter $a$ and effective molecular volume $b$ are dependent on temperature and composition to account for effects of intermolecular forces. For mixtures, the parameters $a$ and $b$ are evaluated as~\cite{poling2001properties}
\begin{subequations}
    \label{EQ_PR_EQUATION_ab}
    \begin{align}
        a &=  \sum^{N_S}_{\alpha=1}\sum^{N_S}_{\beta=1} X_\alpha X_\beta a_{\alpha\beta}\,,\\
        b &= \sum^{N_S}_{\alpha=1} X_\alpha b_\alpha\,,
    \end{align}
\end{subequations}
where $X_\alpha$ is the mole fraction of species $\alpha$. Extended corresponding states principle and single-fluid assumption for mixtures are adopted.~\cite{ely1981prediction, ely1983prediction} The parameters $a_{\alpha\beta}$ and $b_\alpha$ are evaluated using the recommended mixing rules by Harstad et al.~\cite{harstad1997efficient}:
\begin{subequations}
    \label{EQ_PR_EQUATION_alphabeta}
    \begin{align}
        a_{\alpha\beta} &= 0.457236\frac{(R T_{c,\alpha\beta})^2}{p_{c,\alpha\beta}}\left(1+c_{\alpha\beta}\left(1-\sqrt{\frac{T}{T_{c,\alpha\beta}}}\right)\right)^2\;,\\
        b_{\alpha} &= 0.077796 \frac{R T_{c,\alpha}}{p_{c,\alpha}}\;,\\
        c_{\alpha\beta} &= 0.37464 + 1.54226\omega_{\alpha\beta} - 0.26992\omega^2_{\alpha\beta}\;,
    \end{align}
\end{subequations}
where $T_{c,\alpha}$ and $p_{c,\alpha}$ are the critical temperature and pressure of species $\alpha$, respectively. The critical mixture conditions for temperature $T_{c,\alpha\beta}$, pressure $p_{c,\alpha\beta}$, and acentric factor $\omega_{c,\alpha\beta}$ are determined using the corresponding state principles.~\cite{poling2001properties}


\subsection{Thermodynamic properties}
Thermodynamic quantities in this study are evaluated consistently with respect to (w.r.t.) the EoS used and no linearization is introduced.

\subsubsection{Partial derivatives}
Partial derivatives and thermodynamic quantities based on the PR EoS that are useful for evaluating other thermodynamic variables are given as
\begin{subequations}
    \label{eqn:partialderivatives}
    \begin{align}
        \left(\frac{\partial p}{\partial T}\right)_{v, X_i} &= \frac{R}{v-b} - \frac{(\partial a/\partial T)_{X_i}}{v^2+2bv-b^2}\;, \\
        \left(\frac{\partial p}{\partial v}\right)_{T, X_i} &= - \frac{R T}{(v-b)^2}\left\{1-2a\left[R T(v+b)\left(\frac{v^2+2bv-b^2}{v^2-b^2}\right)^2\right]^{-1}\right\}\;, \\
        \left(\frac{\partial a}{\partial T}\right)_{X_i} &= - \frac{1}{T} \sum^{N_S}_{\alpha=1}\sum^{N_S}_{\beta=1} X_\alpha X_\beta a_{\alpha\beta}G_{\alpha\beta}\;,\\
        \left(\frac{\partial^2 a}{\partial T^2}\right)_{X_i} &= 0.457236 \frac{R^2}{2T}\sum^{N_S}_{\alpha=1}\sum^{N_S}_{\beta=1} X_\alpha X_\beta c_{\alpha\beta}(1+c_{\alpha\beta})\frac{T_{c,\alpha\beta}}{p_{c,\alpha\beta}}\sqrt{\frac{T_{c,\alpha\beta}}{T}}\;,\\
        G_{\alpha\beta} &= \frac{c_{\alpha\beta}\sqrt{\frac{T}{T_{c,\alpha\beta}}}}{1+c_{\alpha\beta}\left(1-\sqrt{\frac{T}{T_{c,\alpha\beta}}}\right)}\,,\\
        K_1 &= \int_{+\infty}^v \frac{1}{v^2+2bv-b^2}dv = \frac{1}{\sqrt{8} b}\ln\left(\frac{v+(1-\sqrt{2})b}{v+(1+\sqrt{2})b} \right) \label{eqn:K1}\;.
    \end{align}
\end{subequations}

\subsubsection{Internal energy and enthalpy}
For general real fluids, thermodynamic quantities are typically evaluated from the ideal-gas value plus a departure function that accounts for the deviation from the ideal-gas behavior. The ideal-gas enthalpy, entropy and specific heat can be evaluated from the commonly used NASA polynomials which have a reference temperature of 298~K. The simple mixture-averaged mixing rule is used for ideal-gas mixtures.

The specific internal energy can be written as,
\begin{equation}
    \label{eqn:generalInternalEnergy}
    e(T, \rho, X_i) = e^{\text{ig}}(T, X_i) + \int_0^\rho \left[p - T\left(\frac{\partial p}{\partial T}\right)_{\rho, X_i}\right] \frac{d\rho}{\rho^2}\,,
\end{equation}
where superscript ``ig" indicates the ideal-gas value of the thermodynamic quantity, and \cref{eqn:generalInternalEnergy} can be integrated analytically for PR EoS:
\begin{equation}
    \label{eqn:internalEnergy}
    e = e^{\text{ig}} + K_1\left[a - T\left(\frac{\partial a}{\partial T}\right)_{X_i}\right],
\end{equation}
in which $K_1$ is computed through \cref{eqn:K1}. The specific enthalpy can be evaluated from the thermodynamic relation $h = e + pv$, and we have
\begin{equation}
    h = h^\text{ig} - R T + K_1\left[a - T\left(\frac{\partial a}{\partial T}\right)_{X_i}\right] + pv\,.
\end{equation}

Partial enthalpies for each species, $h_k$,  can be evaluated using the partial derivatives w.r.t. mole fractions. The procedures for evaluating partial enthalpy for real fluids are similar to those in Meng et al.~\cite{meng2003unified} and therefore omitted here.

\subsubsection{Specific heat capacity}
The specific heat capacity at constant volume is evaluated as
\begin{equation}
    c_v = \left(\frac{\partial e}{\partial T}\right)_{v,X_i} = c^\text{ig}_v - K_1 T \left(\frac{\partial^2 a}{\partial T^2}\right)_{X_i}\,,
\end{equation}
and the specific heat capacity at constant pressure is evaluated as
\begin{equation}
    c_p = \left(\frac{\partial h}{\partial T}\right)_{p,X_i} = c^\text{ig}_p - R - K_1 T \left(\frac{\partial^2 a}{\partial T^2}\right)_{X_i} - T \frac{(\partial p/\partial T)^2_{v,X_i}}{(\partial p/\partial v)_{T,X_i}}\,.
\end{equation}

\subsubsection{Speed of sound}
The speed of sound for general real fluids can be evaluated as
\begin{equation}
    \label{eqn:sos}
    c^2 = \left(\frac{\partial p}{\partial \rho}\right)_{s, X_i} = \frac{\gamma}{\rho \kappa_T}\,,
\end{equation}
where $\gamma$ is the specific heat ratio and $\kappa_T$ is the isothermal compressibility, which is defined as
\begin{equation}
    \kappa_T = -\frac{1}{v}\left(\frac{\partial v}{\partial p}\right)_{T, X_i}\,.
\end{equation}

\subsection{Transport properties}\label{sec:transport}
The dynamic viscosity and thermal conductivity are evaluated using Chung's high-pressure method \cite{chung1984, chung1988}. This method is known to produce oscillations in viscosity for multi-species mixtures that consist of species with both positive and negative acentric factors \cite{Hickey2013c, ruiz2015numerical}. To solve this problem, a mass-fraction averaged or mole-fraction averaged viscosity evaluated based on viscosity of each individual species can be used. In this study, the negative acentric factor is set to zero only when evaluating the viscosity so that the anomalies in viscosity can be removed. This approach has similar behavior to the mole-fraction averaged approach via numerical tests.

\section{Transcritical Flamelets}
The large Damk\"ohler number  of the supercritical combustion \cite{Zong2008} supports the use of laminar flamelet-based combustion models. The basic assumption of the flamelet model rests on the fact that the reaction zone remains laminar and the diffusive transport is only important in the direction normal to the flame. By recasting the governing equations as a one-dimensional similarity solution in mixture fraction space, the steady flamelet equations for the species and temperature under unity Lewis number assumption can be written as:
\begin{subequations}
    \begin{align}
        -\frac{\rho \chi}{2}\frac{\partial^{2}Y_{k}}{\partial Z^{2}} &= \dot{\omega}_{k} \;,\\
        -\frac{\rho \chi}{2}\frac{\partial^{2}T}{\partial Z^{2}} &= \frac{\rho \chi}{2 c_p}\frac{\partial c_p}{\partial Z} \frac{\partial T}{\partial Z} + \frac{\dot{\omega}_{T}}{c_p}\;,
    \end{align}
\end{subequations}
where $Y_k$ is the mass fraction of species $k$, $Z$ is the mixture fraction, $\dot{\omega}_k$ is the reaction rate of species~$k$, $\dot{\omega}_{T}$ is the heat release term, and $\chi = 2D |\nabla Z|^2$ is the scalar dissipation rate where $D$ is the diffusion coefficient. No further assumptions are needed to the flamelet equations for a generalized equation of state (for a constant pressure combustion) apart from the modified thermodynamic relationship between temperature and density. For a given profile of the scalar dissipation (which accounts for the convective and diffusive terms normal to the flame front), the solution of the above equations represents the composition and temperature profiles within the counterflow-diffusion flame. In the low-Mach and low-pressure formulation, Peters~\cite{Peters1983} derived a scalar dissipation rate profile for a shear layer under the assumption of a unitary Chapman-Rubesin parameter that relates the local ratio of density and viscosity to its far-field values. Although formally insufficient for many combustion cases, the modeling assumptions of the scalar dissipation rate profile are robust. For transcritical conditions, we retain the same scalar dissipation rate profile. 

\begin{figure}[!b!]
    \centering
    \includegraphics[width=7.2cm,clip=]{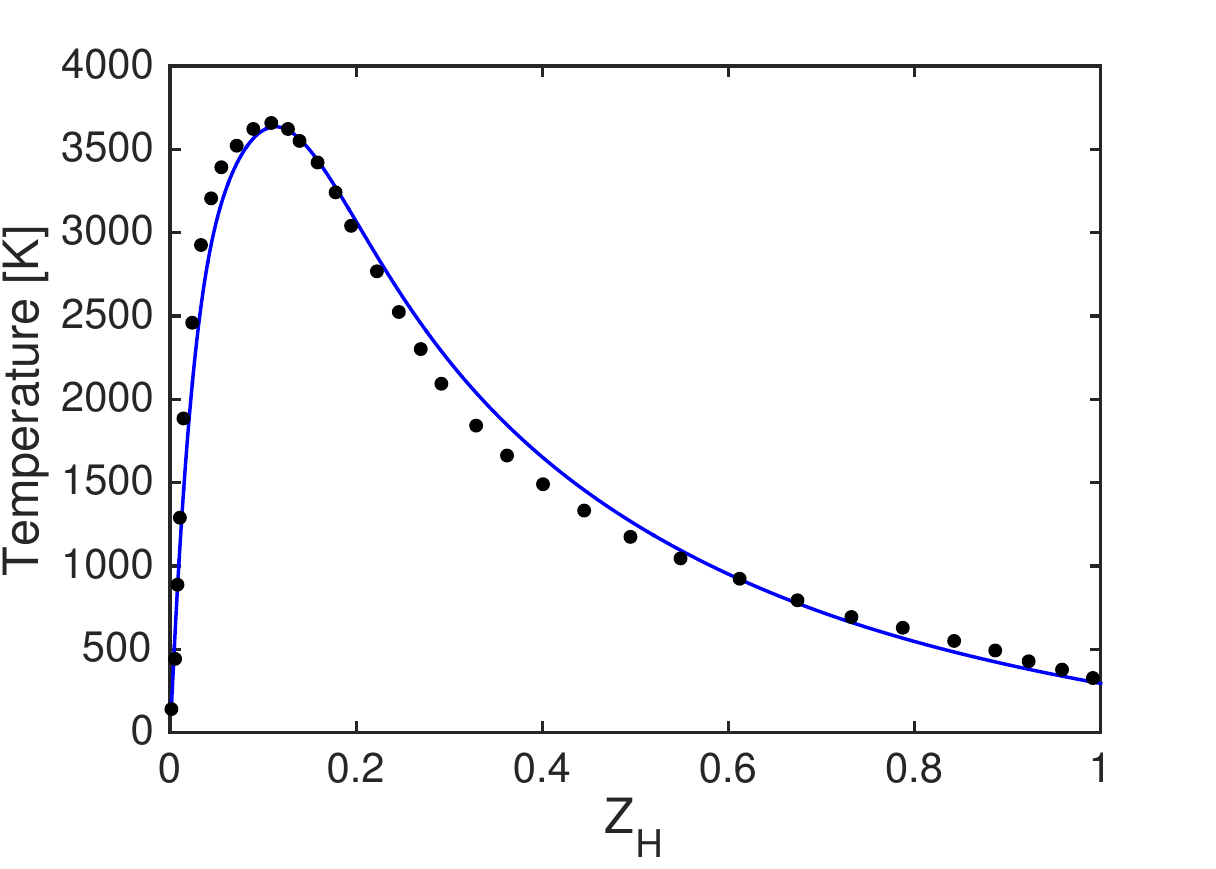}
    \includegraphics[width=7.2cm,clip=]{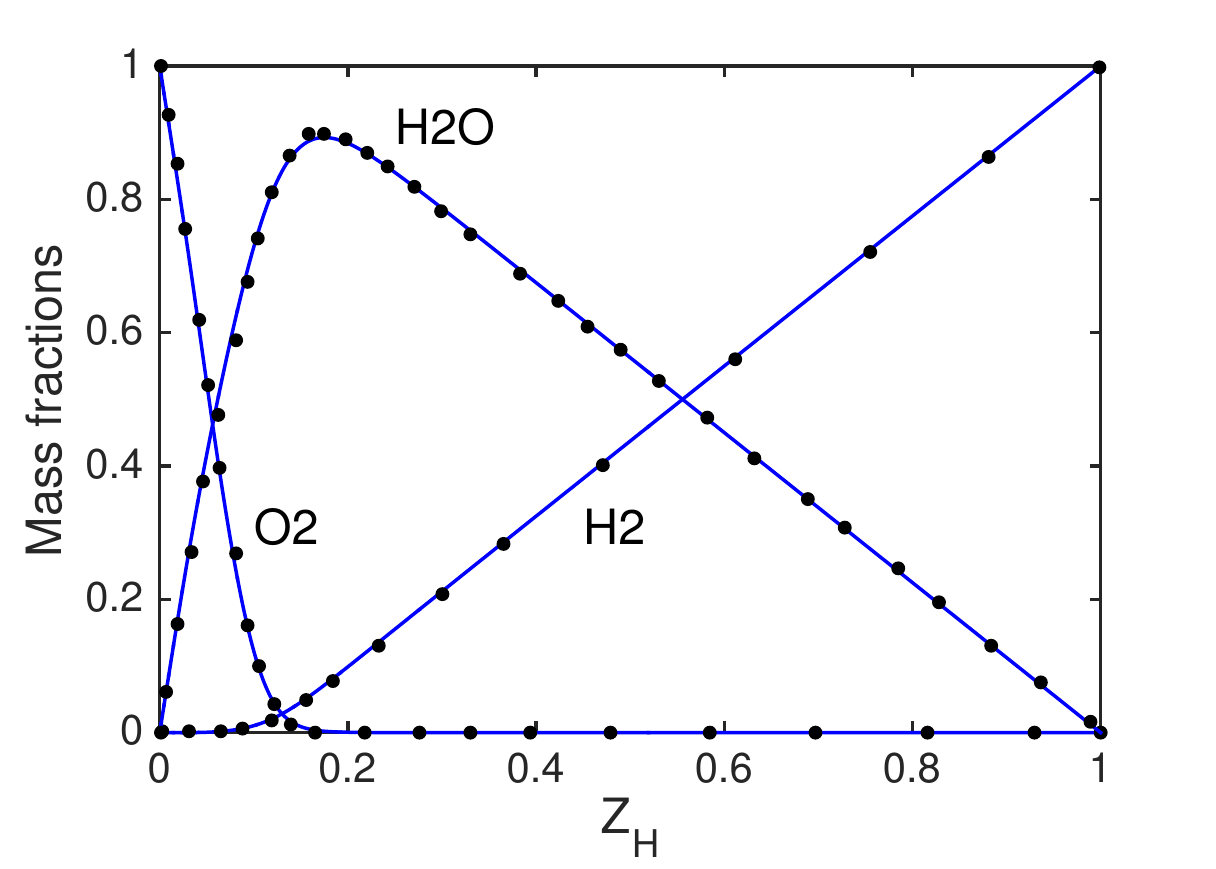}
    \includegraphics[width=7.2cm,clip=]{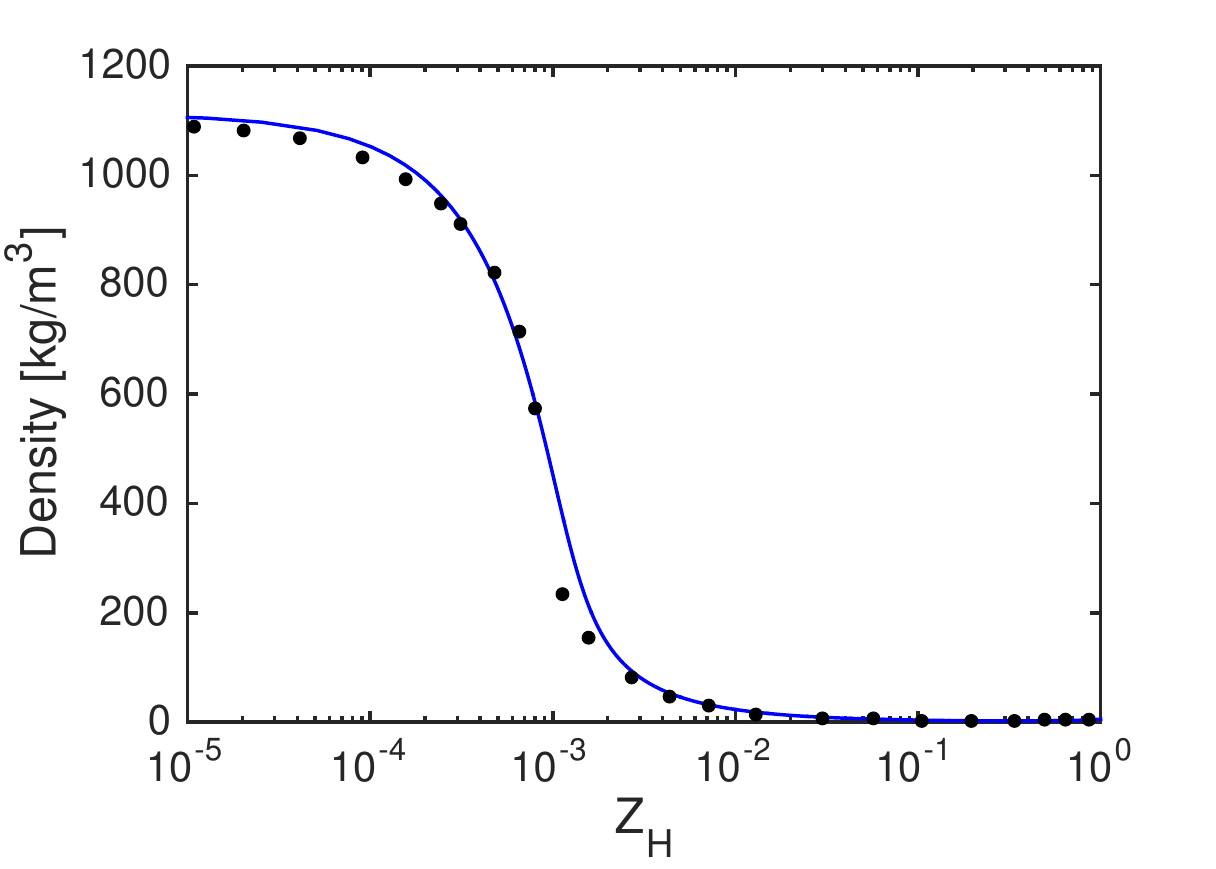}
    \includegraphics[width=7.2cm,clip=]{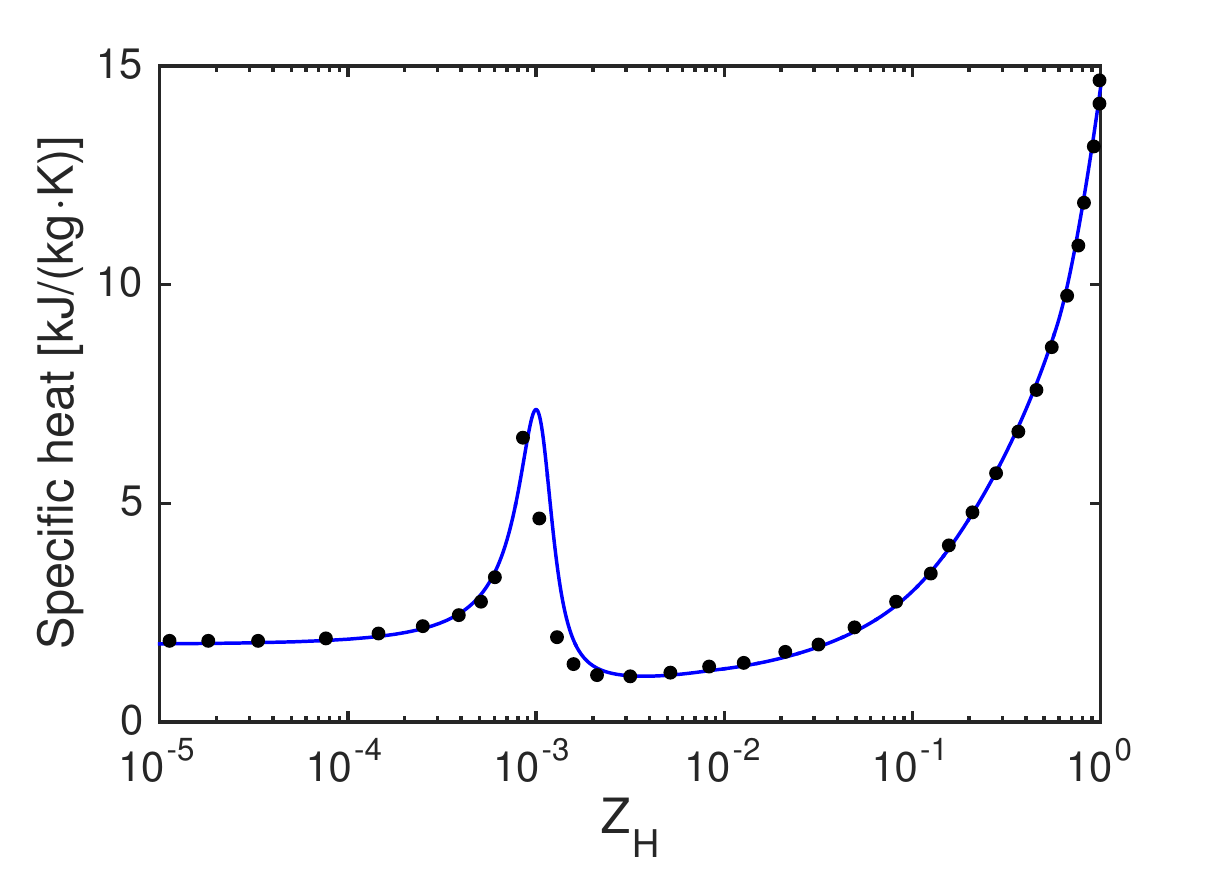}
    \caption{\label{fig:LacazeTandY} Comparisons of temperature, mass fractions, density and specific heat between flamelet (lines) and DNS~\cite{Lacaze2012} (symbols) results for a counterflow diffusion flame with H$_2$ at 295~K and O$_2$ at 120~K at a pressure of 7.0~MPa. }
\end{figure}

The flamelet equations of a counterflow-diffusion flame, in mixture fraction space, are solved using the FlameMaster solver~\cite{Pitsch2006a}. The original code was extended to incorporate the PR EoS, along with the thermodynamically consistent departure functions and thermo-physical fluid properties. The results of the one-dimensional problem are compared with the two-dimensional direct numerical simulations performed by Lacaze and Oefelein \cite{Lacaze2012}. In order to avoid inconsistencies in determining the scalar dissipation profile, only the near-equilibrium flamelet solutions are compared. The setup consists of a counterflow diffusion flame with pure H$_{2}$ and O$_{2}$ streams, respectively, at 295~K and 120~K and a pressure of 7.0~MPa. The strain rate, defined as the velocity difference between both injectors, is set to $10^5$~s$^{-1}$, which corresponds to a scalar dissipation rate of $10^3$~s$^{-1}$, following the classical relationship between the strain and the scalar dissipation rate in laminar flamelets derived by Peters~\cite{Peters1983}.  The high-pressure chemical mechanism by Burke et al.~\cite{Burke2012} is used, which accounts for 8 species and 27 reactions. The comparative results for temperature, composition, density and specific heat capacity are shown in \cref{fig:LacazeTandY}. Specificial attention needs to be paid to the transcritical regions away from the flame front, especifically in the oxidizer stream. In this region, large variations of the thermophysical properties are observed which require an adaptive mesh refinement based on the thermophysical properties in addition to the usual gradient-based mesh adaptation techniques. Without these strong non-linear features, the full complexity of the transcritical behavior cannot be captured. The grid adaptation is especially important to capture the pseudo-boiling point~\cite{Oschwald1999, banuti2015crossing} (PBP) which contains a finite peak in the specific heat capacity, and this region acts as a proxy to phase change in subcritical thermodynamics. Note that the location of the PBP region is typically at mixture fraction around $10^{-3}$ independent of the value of scalar dissipation rate. This requires special attention during the tabulation process since a typical resolution in mixture fraction will completely miss the PBP region, and this will be discussed in details later. As can be seen in \cref{fig:LacazeTandY}, when the thermodynamic features are well resolved, the low-dimensional flamelet equations in mixture fraction space can capture the essential flame structure.

\section{Numerical Methods\label{sec:NumImp}}

\subsection{Governing equations}\label{sec:governing}
The governing equations are the Favre-averaged conservation equations of mass, momentum, total energy, mixture fraction, mixture fraction variance, and progress variable, written as follows:
\begin{subequations}
    \label{eqn:governingEqn}
    \begin{align}
        \frac{\partial \bar{\rho}}{\partial t} + \frac{\partial \bar{\rho} \widetilde{u}_j}{\partial x_j} &= 0\,,\\
        \frac{\partial \bar{\rho} \widetilde{u}_i}{\partial t} + \frac{\bar{\rho} \widetilde{u}_i \widetilde{u}_j}{\partial x_j} &= -\frac{\partial \bar{p}}{\partial x_i} + \frac{\partial}{\partial x_j} \left[ (\widetilde{\mu} + \mu_t) \left( \frac{\partial \widetilde{u}_i}{\partial x_j} + \frac{\partial \widetilde{u}_j}{\partial x_i} - \frac{2}{3} \delta_{ij} \frac{\partial \widetilde{u}_k}{\partial x_k} \right) \right] \,,\label{eqn:momentumequation} \\
        \frac{\partial  \bar{\rho} \widetilde{E}}{\partial t} + \frac{\bar{\rho} \widetilde{u}_j \widetilde{E}}{\partial x_j} &=
        \begin{aligned}[t]
            &\frac{\partial}{\partial x_j} \left[ \left(\widetilde{\frac{\lambda}{c_p}} + \frac{\mu_t}{\text{Pr}_t} \right) \frac{\partial \widetilde{h}}{\partial x_j} - \widetilde{u}_j \bar{p} + \widetilde{u}_i (\bar{\tau}_{ij} + \bar{\tau}^R_{ij}) \right] \\
            &+ \frac{\partial}{\partial x_j} \left[ \sum_{k = 1}^N \left(\bar{\rho}\widetilde{D}_k - \widetilde{\frac{\lambda}{c_p}} \right) \widetilde{h}_k \frac{\partial \widetilde{Y}_k}{\partial x_j} \right] \,,
        \end{aligned} \label{eqn:energyequation} \\
        \frac{\partial \bar{\rho} \widetilde{Z}}{\partial t} + \frac{\bar{\rho} \widetilde{u}_j \widetilde{Z}}{\partial x_j} &= \frac{\partial}{\partial x_j} \left[ \left( \bar{\rho}\widetilde{D} + \frac{\mu_t}{\text{Sc}_t} \right) \frac{\partial \widetilde{Z}}{\partial x_j} \right]\,,\\
        \frac{\partial \bar{\rho} \widetilde{Z^{''2}}}{\partial t} + \frac{\bar{\rho} \widetilde{u}_j \widetilde{Z^{''2}}}{\partial x_j} &= \frac{\partial}{\partial x_j} \left[ \left( \bar{\rho}\widetilde{D} + \frac{\mu_t}{\text{Sc}_t} \right) \frac{\partial \widetilde{Z^{''2}}}{\partial x_j} \right] + 2\frac{\mu_t}{\text{Sc}_t}\frac{\partial \widetilde{Z}}{\partial x_j}\frac{\partial \widetilde{Z}}{\partial x_j} - \bar{\rho} \widetilde{\chi}\,,\\
        \frac{\partial \bar{\rho} \widetilde{C}}{\partial t} + \frac{\bar{\rho} \widetilde{u}_j \widetilde{C}}{\partial x_j} &= \frac{\partial}{\partial x_j} \left[ \left( \bar{\rho}\widetilde{D} + \frac{\mu_t}{\text{Sc}_t} \right) \frac{\partial \widetilde{C}}{\partial x_j} \right] + \bar{\dot{\omega}}_C\,,
    \end{align}
\end{subequations}
where $u_i$ is the $i$th component of the velocity vector, $E$ is the total energy including the chemical energy, $C$ is the progress variable, $\mu$ and $\mu_t$ are the laminar and turbulent viscosity, $\lambda$ is the thermal conductivity, $D$ is the diffusion coefficient for the scalars, $\dot{\omega}_C$ is the source term for the progress variable, $\tau_{ij}$ and $\tau_{ij}^R$ are the viscous and subgrid-scale stresses which are assumed to take the form as the second term on the right-hand side of \cref{eqn:momentumequation}, Pr$_t$ is the turbulent Prandtl number, and Sc$_t$ is the turbulent Schmidt number. An appropriate subgrid-scale model is needed for the computation of the turbulent viscosity $\mu_t$. Under the unity Lewis number assumption, the summation on the right-hand side of \cref{eqn:energyequation} vanishes so that the species mass fractions are not explicitly required for the energy equation. The system is closed with the PR EoS introduced in \cref{sec:RealFluid} and the flamelet-based combustion model that will be discussed in the next subsection. Moreover, the sub-grid terms associated with the EoS are neglected in this study.

\subsection{Combustion model\label{sec:combustionmodel}}
A Flamelet/Progress Variable (FPV) approach~\cite{Pierce2004, Ihme2005} is adopted in this study, in which the chemistry is pre-computed and tabulated as a series of laminar flamelet solutions. Flamelets are first computed for different values of the scalar dissipation rate at a constant background pressure and specified constant fuel and air temperatures, and then the flamelets are parametrized by the mixture fraction and the reaction progress variable. The resulting flamelet table is used for the determination of the local temperature, species, density, source term of the progress variable and other thermal-transport quantities needed by the solver. Presumed PDFs are introduced to account for the turbulence/chemistry interaction. Typically, a $\beta$-PDF is used for the mixture fraction and a $\delta$-PDF for the progress variable, which was shown to be a reasonable approximation in many studies. Since the reacting region is typically in the ideal gas regime even under transcritical combustion conditions, the PDF closures are expected to perform similarly as in previous studies for ideal gas reacting flows.

In the low-Mach number flamelet implementation, the temperature, species, and density are assumed to depend only on the transported scalars. However, when compressibility effects are taken into account, an overdetermined thermodynamic state arises from the use of the flamelet table with tabulated thermodynamics. On the one hand, the full thermodynamic state, at constant pressure, is defined within the flamelet table. On the other hand, the transport equations contain two thermodynamic variables, namely density and internal energy, which are also sufficient to fully characterize the thermodynamic equilibrium state of the fluid if the compositions are given. In order to mend the over-determined thermodynamic states, a strategy was developed by Saghafian~et~al.~\cite{saghafian2015efficient} in the context of ideal gas flows to account for the pressure and temperature variations arising in supersonic combustion using the FPV approach. The specific heat ratio is linearized around temperature to eliminate the costly iterative procedure to determine temperature, and also to obtain other thermodynamic quantities which are functions of temperature.

The proposed strategy for applying compressibility effects in the FPV approach has been modified to work with a generalized equation of state under transcritical conditions in the present work. The underlying strategy rests on correcting the tabulated values with the transported quantities based on the EoS used. Specifically, since PR EoS is used in this study, along with thermodynamic quantities needed for evaluation of the ideal gas thermodynamic quantities, parameters $a$, $b$, and the first and second derivatives of the parameter $a$ w.r.t. temperature are needed for the calculations of the partial derivatives in \cref{eqn:partialderivatives} that are needed for the evaluation of the departure functions. The parameter $b$ is a function of species composition only and is independent of pressure and temperature. Therefore, it can be pre-tabulated within the flamelet table without any discrepancy in pressure and temperature. However, the parameter $a$, along with its derivatives, is a function of both the species composition and the temperature, and thus may not be consistent with the temperature corresponding to the transported variables. The following procedure is proposed for the evaluation of the parameter $a$ and its derivatives: the dependence of the parameter $a$ on temperature is assumed to be a quadratic function as follows,
\begin{equation}
    a = C_1 \widetilde{T}^2 + C_2 \widetilde{T} + C_3\,, \label{eqn:aT}
\end{equation}
where $C_1$, $C_2$, and $C_3$ can be determined from tabulated quantities,
\begin{subequations}
    \label{eqn:C1C2C3}
    \begin{align}
        C_1 &= \frac{1}{2} \left(\frac{\partial^2 a}{\partial T^2}\right)_0 \;, \\
        C_2 &= \left(\frac{\partial a}{\partial T}\right)_0 - 2 C_1 T_0\;, \\
        C_3 &= a_0 - C_1 {T_0}^2 - C_2 T_0\;,
    \end{align}
\end{subequations}
where subscript 0 indicates the stored baseline quantities in the table. The first and second derivatives of parameter $a$ w.r.t. temperature can be determined accordingly by taking derivatives of \cref{eqn:aT}, and the proposed model corresponds to a linear and a constant approximation to the first and second derivatives, respectively. Once the parameter $a$ and its derivatives are obtained, along with the parameter $b$ and the gas constant $R$, all the partial derivatives which are needed for computing other thermodynamic quantities can be evaluated for a given mixture, and therefore, the thermodynamic state is determined. Note that similar to the quadratic model as in \cref{eqn:aT}, a linear model or a constant model for the parameter $a$ can also be constructed based on the stored values in the table. The performance of the quadratic, linear, and constant model will be examined later.

\begin{figure}[!tb!]
    \centering
    \includegraphics[width=76mm]{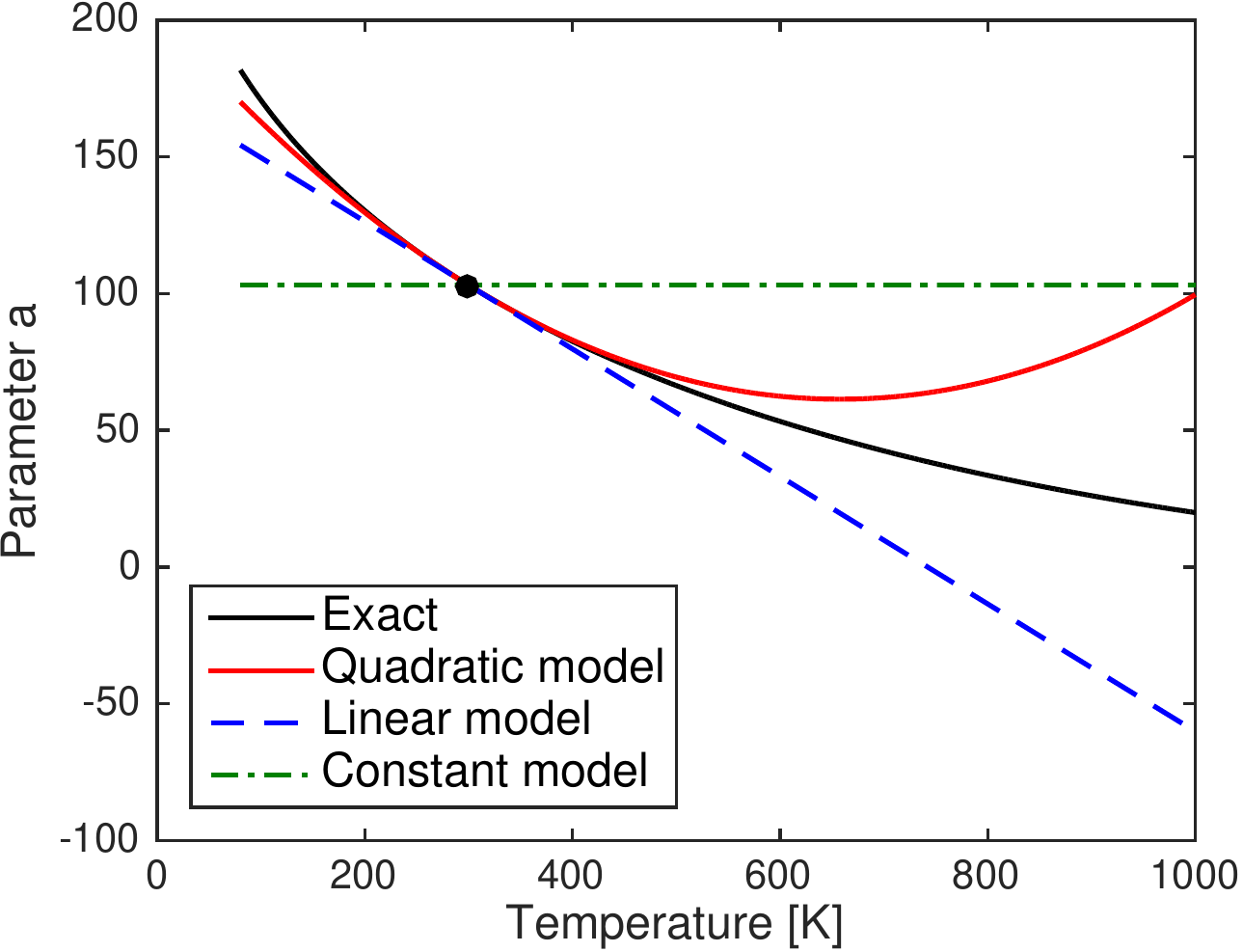} \quad
    \includegraphics[width=76mm]{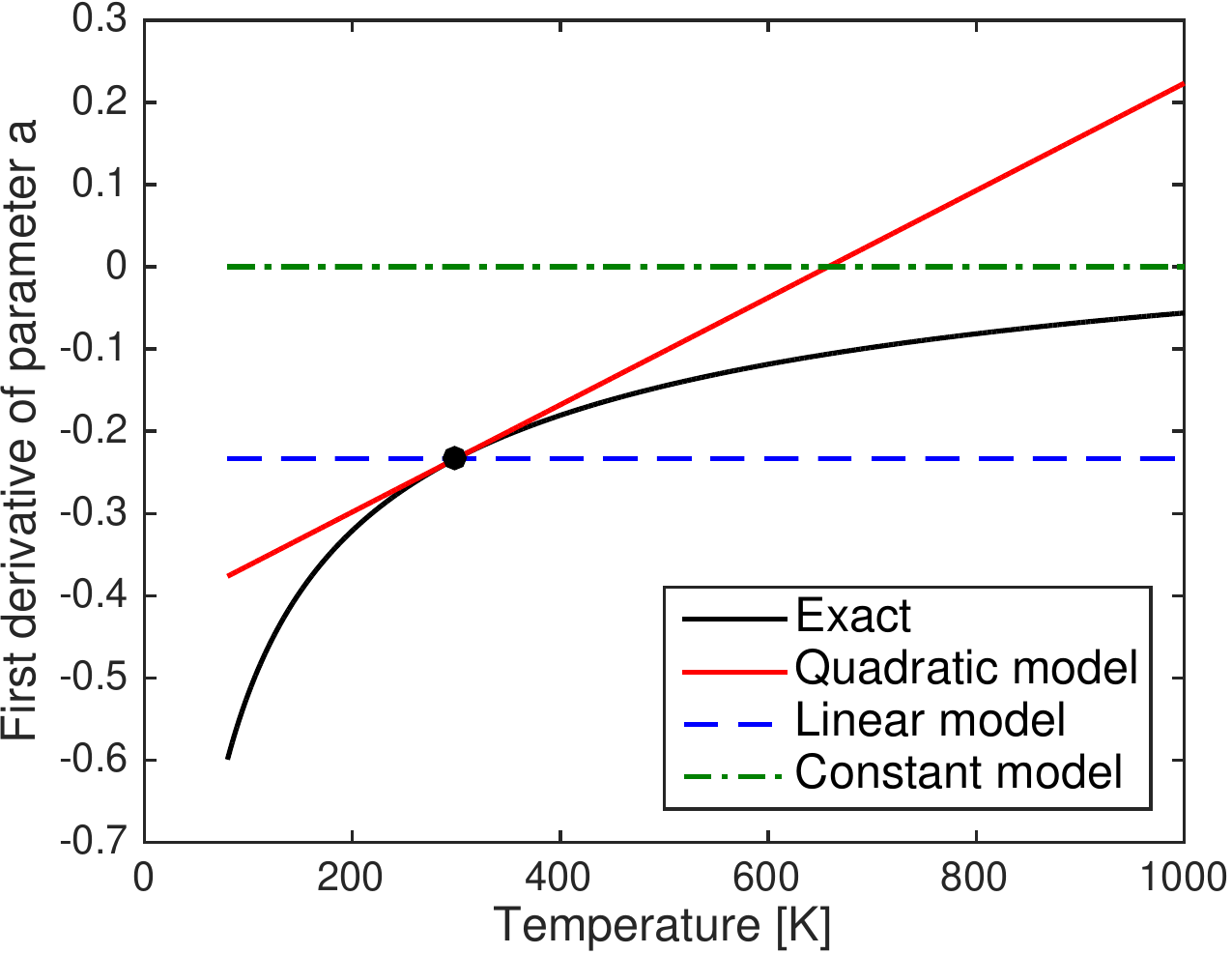}
    \caption{Illustration of the quadratic, linear and constant model for parameter $a$. Results for parameter $a$ (left) and its first derivative w.r.t. temperature (right) are shown. Exact solution is for O$_2$ at 100 bar. Black dot indicates the reference point at 300 K.\label{fig:examplea}}
\end{figure}

As an illustration, \cref{fig:examplea} shows the results of parameter $a$ and its first derivative w.r.t. temperature, along with the approximations evaluated based on the quadratic, linear, and constant model. Pure oxygen at 100~bar is considered for this example. The reference point is assumed to be at a temperature of 300~K, as indicated by the black dot in \cref{fig:examplea}. As can be seen from \cref{fig:examplea}, the quadratic assumption works well in the region within 200~K and 100~K from the reference temperature for the parameter $a$ and its derivative, respectively. The linear model yields a linear profile for $a$ and hence a constant profile for its derivative. The constant model for the parameter $a$ gives zero value for its derivative. Similarly, the behavior of the second derivative of $a$, which is also important for the evaluation of real-fluid thermodynamics, can be expected for the three models considered. The quadratic model for $a$ shows superior performance in predicting $a$ and its derivatives when temperature is away from the reference value, and its performance in predicting other thermodynamic quantities will be examined in details to confirm the validity of the proposed approach.

As an example, to calculate the internal energy including the chemical energy from the temperature based on the proposed approach for PR EoS for a given mixture, i.e. fixed $\widetilde{Z}$, $\widetilde{Z^{''2}}$, and $\widetilde{C}$, the ideal gas part and the departure function are calculated separately,
\begin{equation}
    \widetilde{e} = \widetilde{e}^{\,\text{ig}} + \widetilde{e}^{\,\text{dep}}\,,
\end{equation}
where $\widetilde{e}^{\,\text{ig}}$ and $\widetilde{e}^{\,\text{dep}}$ are the ideal-gas and departure function values of the specific internal energy. The ideal-gas value including the chemical energy of the mixture is calculated with linearized specific heat ratio~\cite{saghafian2015efficient}
\begin{equation}
    \widetilde{e}^{\,\text{ig}} = \widetilde{e}_0^{\,\text{ig}} + \frac{\widetilde{R}}{a_\gamma^\text{ig}} \text{ln} \left( 1 + \frac{a_\gamma^\text{ig} (\widetilde{T} - T_0)}{\widetilde{\gamma}_0^{\,\text{ig}} - 1} \right) \,,
\end{equation}
where $\widetilde{e}_0^{\,\text{ig}}$, $\widetilde{R}$, $T_0$, $\widetilde{\gamma}_0^{\,\text{ig}}$, and $a_\gamma^\text{ig}$ are all stored variables in the table for given $\widetilde{Z}$, $\widetilde{Z^{''2}}$, and $\widetilde{C}$, in which $a_\gamma^\text{ig}$ is the slope of the ideal-gas specific heat ratio w.r.t. the temperature. The departure function is determined as
\begin{equation}
    \widetilde{e}^{\,\text{dep}} = K_1 \left[a - \widetilde{T}\left(\frac{\partial a}{\partial T}\right)_{X_i}\right]\,,
\end{equation}
where \cref{eqn:aT,eqn:C1C2C3} are used for the evaluation of parameters needed for the PR EoS. Other thermodynamic quantities, such as the specific heat and speed of sound, can be evaluated similarly.

To determine primitive variables from conservative variables, a secant method is used to obtain temperature given the transported density and internal energy. With this, the pressure along with other thermodynamic quantities is evaluated as an explicit function of the density and temperature. Since the parameters needed for the real fluid EoS are pre-tabulated and approximated from the table, the computational overhead of the iteration process is acceptable.

Transport quantities are evaluated based on the method due to Chung et al.~\cite{chung1984, chung1988} Since the variation of the viscosity and conductivity on pressure is small under transcritical conditions, a power-law is used to approximate the temperature dependency as follows,
\begin{subequations}
    \begin{align}
        \frac{\widetilde{\mu}}{\widetilde{\mu}_0} &= \left(\frac{\widetilde{T}}{T_0} \right)^{a_\mu} \;, \\
        \frac{\widetilde{\lambda}}{\widetilde{\lambda}_0} &= \left(\frac{\widetilde{T}}{T_0} \right)^{a_\lambda}\;\,,
    \end{align}
\end{subequations}
where $\widetilde{\mu}_0$ and $\widetilde{\lambda}_0$ are stored in the table along with their corresponding slopes, $a_\mu$ and $a_\lambda$.

Note that the proposed approach is not limited to the PR EoS. All cubic EoS's have similar structure and a similar approach can be used for other types of cubic EoS, such as the Soave-Redlich-Kwong (SRK) EoS~\cite{soave1972equilibrium}. 

The developed FPV approach focuses on the thermodynamics and no special modification is taken for the chemistry part. The transcritical combustion dynamics were shown to have similar structures as for ideal gases by several studies~\cite{Ribert2008, huo2014general, banuti2016sub}. Indeed, in typical engines, combustion takes place at high temperatures away from the real-fluid region which is characterized by cryogenic temperatures. Moreover, for the applications considered in this study, combustion can be considered to be at low-Mach conditions where compressibility effects can be neglected. However, for the inert and equilibrium parts of the flows without chemical source terms, compressibility may play a critical role for predicting the behaviors of the system of interest, for example flows in the injectors and through the nozzles for rocket engines. Therefore, a strategy is proposed here for practical simulations, that the flamelet table can be generated at conditions of the combustion chamber where combustion takes place, and discrepancies in temperature and pressure in other parts of the flows can be corrected by the FPV model described above. If supersonic combustion needs to be considered, methodologies developed by Saghafian~et~al.~\cite{saghafian2015efficient} can be used.

\subsection{Numerical schemes}
The massively paralleled, finite-volume solver, {\it CharLES$^{\,x}$}, developed at the Center for Turbulence Research, is used in this study. A control-volume based finite volume approach is utilized for the discretization of the system of equations, \cref{eqn:governingEqn}:
\begin{equation}
    \frac{\partial U}{\partial t} V_{cv} + \sum_f F^e A_f = \sum_f F^v A_f + SV_{cv}\,,
\end{equation}
where $U$ is the vector of conserved variables, $F^e$ is the face-normal Euler flux vector, $F^v$ is the face-normal viscous flux vector which corresponds to the r.h.s of \cref{eqn:governingEqn}, $S$ is the source term vector, $V_{cv}$ is the volume of the control volume, and $A_f$ is the face area. A strong stability preserving 3rd-order Runge-Kutta (SSP-RK3) scheme \cite{gottlieb2001strong} is used for time advancement.

The convective flux is discretized using a sensor-based hybrid scheme in which a high-order, non-dissipative scheme is combined with a low-order, dissipative scheme to minimize the numerical dissipation introduced. A central scheme which is fourth-order on uniform meshes is used along with a second-order ENO scheme for the hybrid scheme. A density sensor~\cite{Hickey2013c, ma2016entropy} is adopted in this study. Due to the large density gradients across the PBP region under transcritical conditions, an entropy-stable flux correction technique, developed by Ma et al.~\cite{ma2016entropy}, is used to ensure the physical realizability of the numerical solutions and to dampen the non-linear instabilities in the numerical schemes. 

Due to the strong non-linearlity inherited in the real fluid EoS, spurious pressure oscillations will be generated when a fully conservative scheme is used~\cite{terashima2012approach, ma2016entropy}, and severe oscillations in the pressure field could make the solver diverge which cannot be solved by adding artificial dissipation. A double-flux method~\cite{ma2014supercritical, ma2016numerical, ma2016entropy} is extended to the transcritical regime to eliminate the spurious pressure oscillations. The effective specific heat ratio based on the speed of sound is frozen both spatially and temporally for a given cell when the fluxes of its faces are evaluated, which renders a local system as an equivalently ideal-gas system. The two fluxes at a face evaluated in the double-flux method are not the same, yielding a quasi-conservative scheme and the conservative error in total energy was shown to converge to zero with increasing resolution.~\cite{ma2016entropy} A Strang-splitting scheme~\cite{strang1968construction} is applied in this study to separate the convection operator from the remaining operators of the system.

\section{Results and Discussions}

\subsection{Tabulation approach analysis}
The validity and performance of the FPV model with the tabulation approach developed in the previous sections will be assessed in the following through an {\it a priori} analysis. To this end, a flamelet table is firstly generated at certain reference conditions and then a flamelet solution at different reference conditions is used as a probe for the assessment. In a compressible solver, initial and boundary conditions are typically specified through primitive variables (temperature, pressure, velocity, and species). With the FPV approach, species are determined from mixture fraction and progress variable. The solver then converts primitive variables to conservative variables and takes time advancement. To be consistent with the compressible solver, the primitive variables, namely temperature, pressure, mixture fraction and progress variable, from the probing flamelet solution are given, and thermodynamic parameters, such as gas constant, specific heat ratio, parameter $a$ and $b$ for the PR EoS, are interpolated from the table by fixing mixture fraction and progress variable. Then the species and thermodynamic quantities of interest are compared to the exact values from the probing flamelet to assess the performance of the current developed tabulation approach.

\begin{figure}[!t!]
    \centering
    \includegraphics[width=14.2cm,clip=]{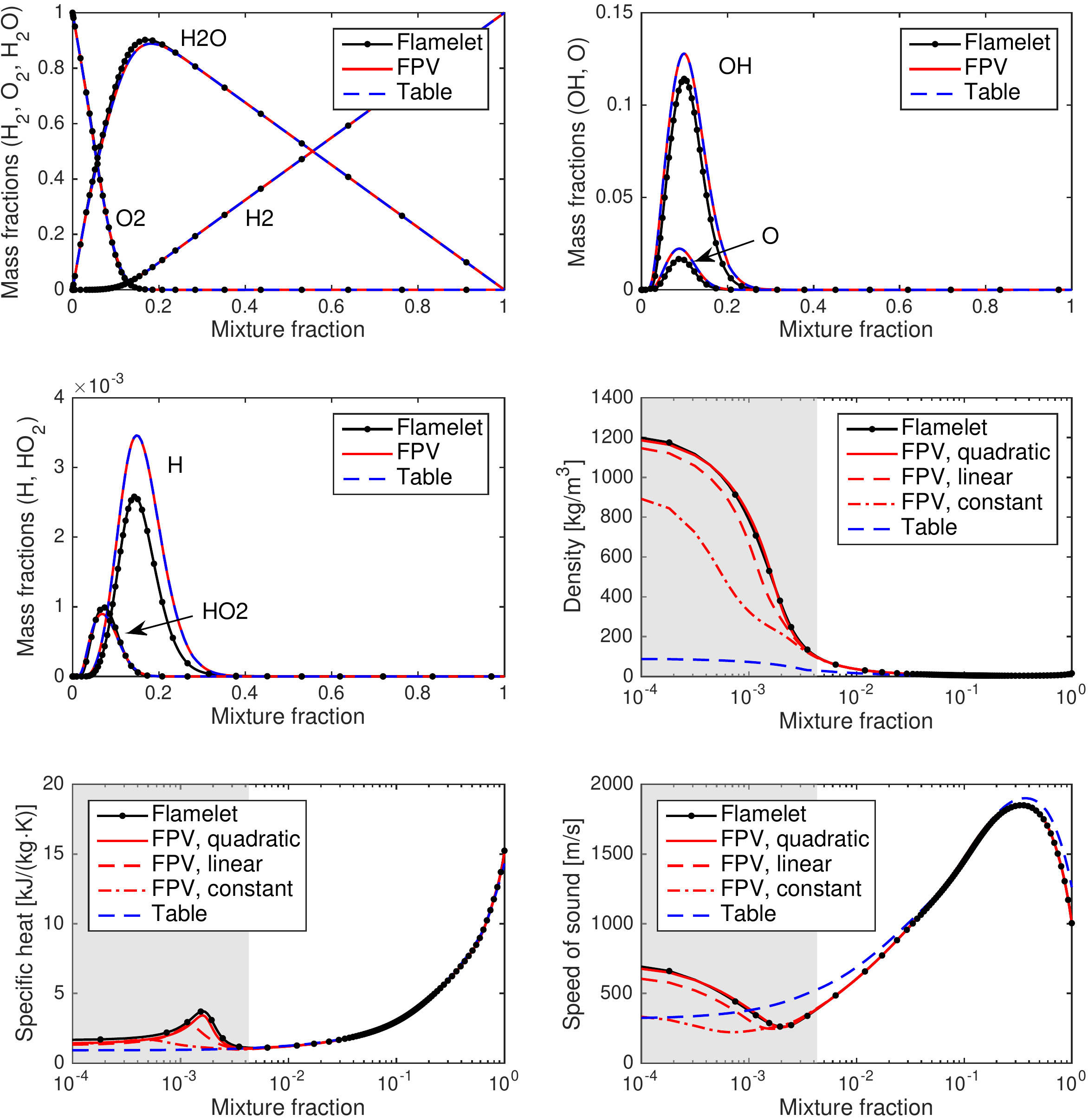}
    \caption{\label{fig:analysis} Mass fractions (H$_2$, O$_2$, H$_2$O, OH, O, H, HO$_2$), density, specific heat and speed of sound predicted by the current FPV approach with quadratic, linear, and constant models for parameter $a$ in comparison with exact values from a probing flamelet under transcritical conditions ($T_\text{ox}$~=~100~K, $T_\text{f}$~=~150~K, $p$~=~100~bar). Table used by the FPV approach is constructed using flamelets under ideal-gas conditions at a different pressure ($T_\text{ox}$~=~300~K, $T_\text{f}$~=~300~K, $p$~=~60~bar). Temperature, pressure, mixture fraction and progress variables in the FPV approach are obtained from the probing flamelet. Black dotted line represents the exact values from the probing flamelet. Red lines represents predictions from the current FPV approach. Blue dashed line corresponds to values stored in the table. Shaded area indicates real-fluid region.}
\end{figure}

An extreme case is considered here to show the capabilities of the current model to recover real-fluid thermodynamics. The flamelet table is generated at ideal-gas conditions with $T_\text{ox}$~=~300~K, $T_\text{f}$~=~300~K, $p$~=~60~bar. A transcritical flamelet in equilibrium at relatively low scalar dissipation rate at conditions $T_\text{ox}$~=~100~K, $T_\text{f}$~=~150~K, $p$~=~100~bar is considered as the probing flamelet. Note that the flamelet table contains only ideal-gas information and directly reading variables from the table is expected to give significant errors. This situation corresponds to the case when an ideal-gas table is used for transcritical simulations, or when the table resolution in mixture fraction is insufficient, so that the PBP region is not resolved.

\Cref{fig:analysis} shows the results of the {\it a priori} analysis. Species and thermodynamic quantities of interest evaluated from the current FPV model and directly read from the table are compared to the exact values from the probing flamelet. The current FPV model assumes that the composition is not changed with perturbations on the reference conditions, or in other words, species are only functions of the mixture fraction and the progress variable. Thus species from the FPV model are expected to have the same values as in the flamelet table. Quadratic, linear, and constant models for the parameter $a$ are compared in terms of real-fluid thermodynamic quantities. In the following, we will go through different aspects contained in this analysis.

The species results, including major species (H$_2$, O$_2$, and H$_2$O) and representative minor species (OH, O, H, and HO$_2$), are shown in the first three subfigures in \cref{fig:analysis}. It can be seen that the major species read from the table exhibit negligible difference from the exact values even considering the dramatic difference in the pressure. For minor species, small but noticeable discrepancies can be observed, but the influence on the mixture properties is expected to be insignificant due to the low magnitude of their mass fractions. These results are consistent with the findings by Saghafian~et~al.~\cite{saghafian2015efficient} where flamelets at different operating conditions are extensively studied for ideal gases at pressures close to ambient conditions. Similar results are found for the entire S-shaped curve with probing flamelets at different scalar dissipation rates which are not shown here.

\begin{figure}[!tb!]
    \centering
    \includegraphics[height=58mm]{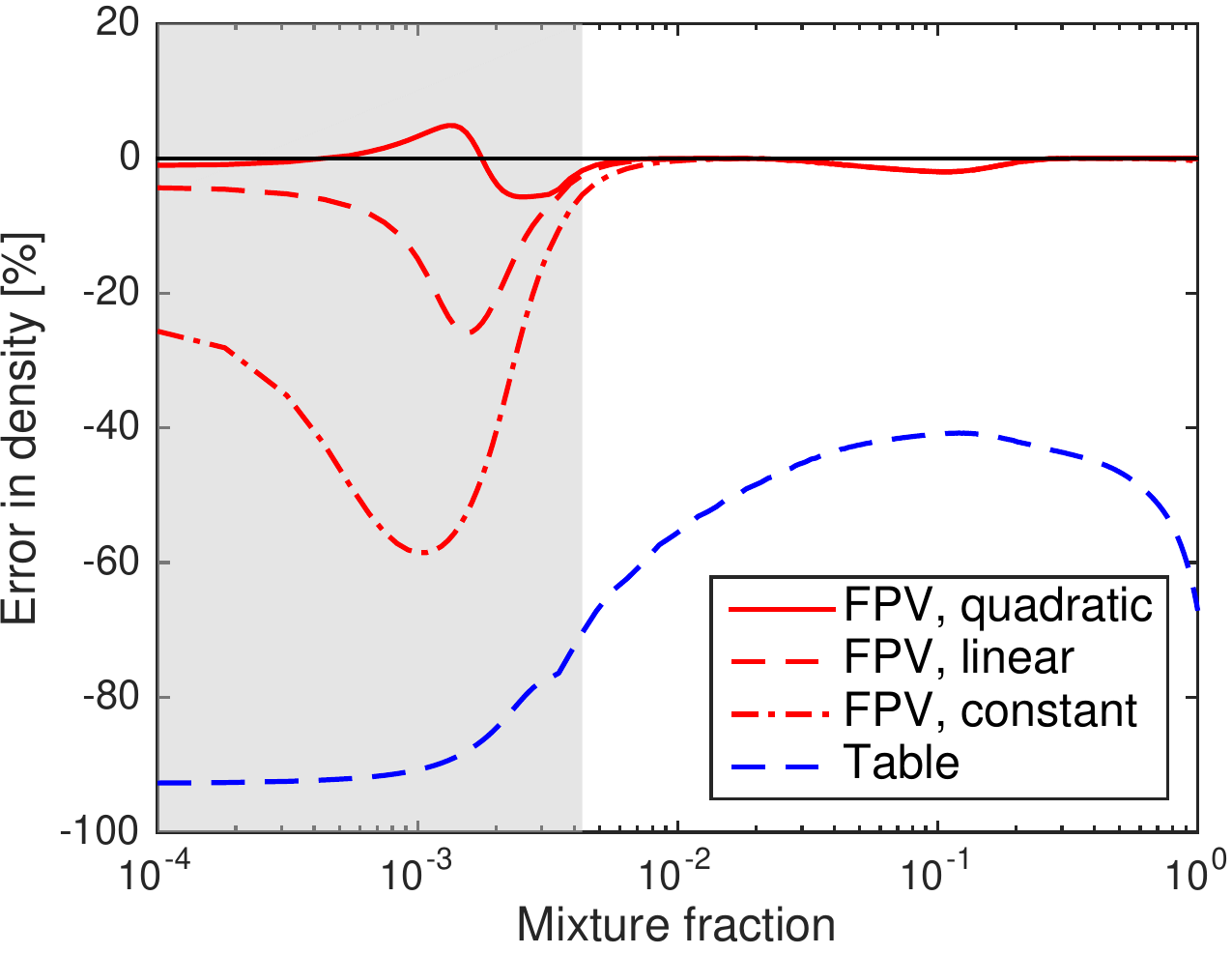} \quad
    \includegraphics[height=58mm]{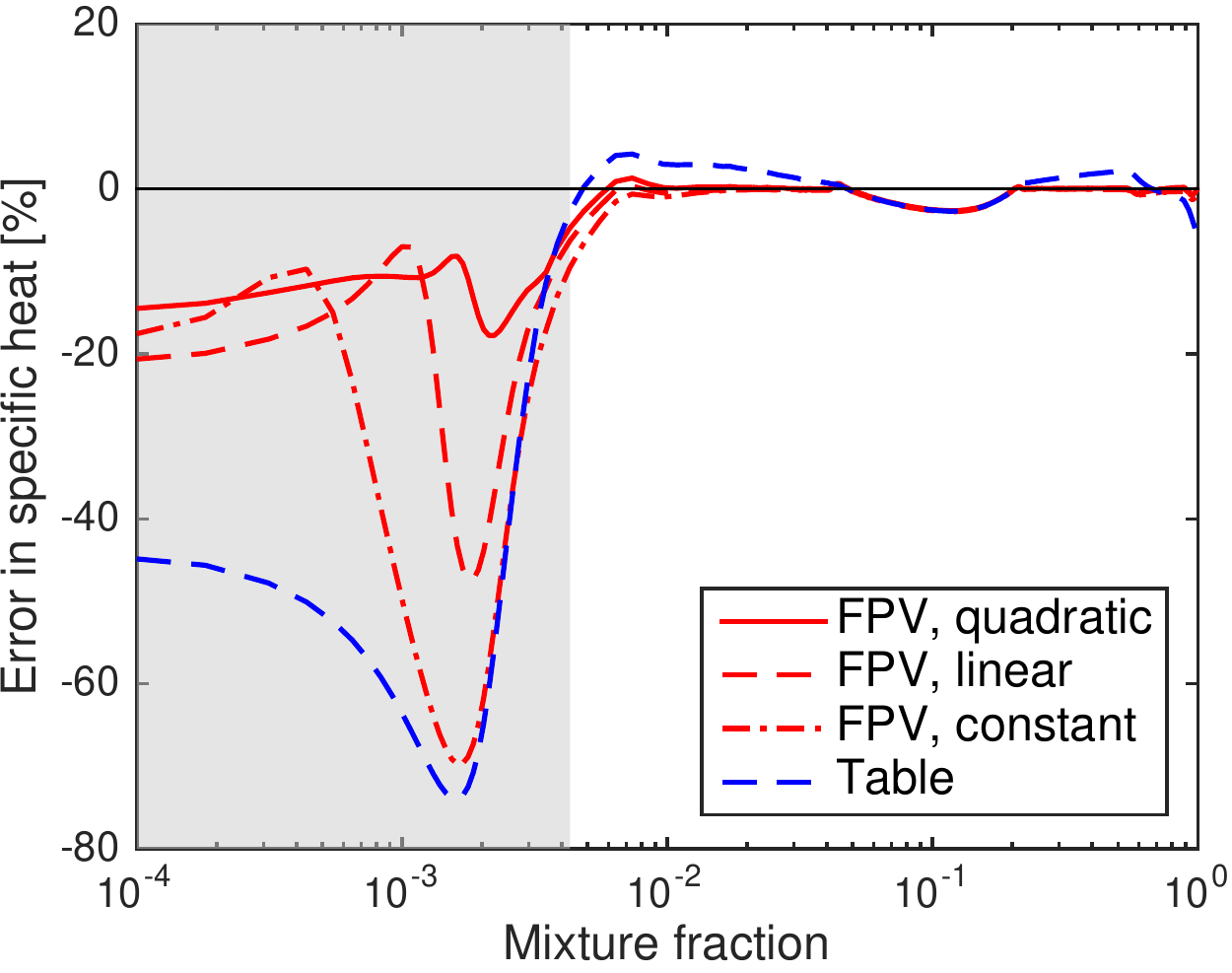}
    \caption{Relative error of density (left) and specific heat (right) in percentage of \cref{fig:analysis}. Shaded area indicates real-fluid region and white area ideal-gas region. \label{fig:analysiserror}}
\end{figure}

Density, specific heat at constant pressure, and speed of sound are used as examples for comparison of the thermodynamic quantities in \cref{fig:analysis}. Since the flamelet table is generated at different reference conditions from the probing flamelet, the quantities directly read from the table show significant discrepancies from the flamelet. The table is at ideal-gas conditions, and therefore, the large density value of LOX and the sharp change of density profile in the PBP region are completely missed. Similarly, the peak in specific heat in the PBP region cannot be read from the table. The behavior in speed of sound is not correct as compared to the probing flamelet in the real-fluid region as indicated by the shaded area. In the ideal-gas region, which is indicated by the white area in the last subfigures of \cref{fig:analysis} (defined by 5\% deviation of compressibility factor from unity), although the gas constant is well recovered by the species information, due to the drastic difference in temperature and pressure, the table gives significant errors especially in density. This is better depicted in \cref{fig:analysiserror}, where relative errors in density and specific heat from different models are shown. As can be seen in \cref{fig:analysiserror}, the density directly interpolated from the table has errors exceeding 80\% and 40\% in the real-fluid and ideal-gas regions, respectively. 

In contrast, the current FPV model with a quadratic model for the parameter $a$ shows superior performance in recovering these quantities. In the ideal-gas region, a linearized specific heat ratio is used to compensate the temperature discrepancy from the table, and this has been shown to yield good predictions with a temperature discrepancy up to 500~K~\cite{pecnik2012reynolds}. Similar behavior can be seen in the current study. As shown in \cref{fig:analysiserror} that the relative error from the FPV model in the ideal-gas region is less than 5\%. In the real-fluid region, despite the fact that extrapolation from the table is needed for recovering the thermodynamic quantities (200~K difference of temperature for the oxidizer stream), the current FPV model shows significant improvements. The quadratic, linear and constant model yield increasingly more accurate results successively towards the flamelet solution as expected. The relative error of the quadratic model is below 10\% and 20\% for the density and specific heat as shown in \cref{fig:analysiserror}. Obviously when the reference table contains real-fluid information by lowering the oxidizer temperature, and with grid points in mixture fraction clustered in the oxidizer side, the performance of the current FPV model can be further improved. The operating conditions are deliberately chosen to challenge the current FPV model.

In summary, the assumption that the composition is a weak function of the reference conditions is valid under transcritical conditions. The currently developed FPV approach with quadratic expression for the attraction parameter can accurately recover the real-fluid and ideal-gas thermodynamic quantities despite variations in temperature and pressure.

\subsection{Cryogenic LOX/GH2 mixing case}
The proposed FPV approach is tested with a benchmark case proposed by Ruiz~et~al.~\cite{ruiz2015numerical} for high-Reynolds number turbulent flows with large density ratios. A two-dimensional mixing layer of liquid-oxygen (LOX) and gaseous-hydrogen (GH2) streams is simulated. Details of the configuration and the computation domain for the simulation can be found in Ruiz~et~al.~\cite{ruiz2015numerical} The configuration is representative of a coaxial rocket combustor, in which dense LOX is injected in the center to mix with the surrounding high-speed GH2 stream. The two streams are separated by the injector lip which is also included in the computational domain. The computational domain has a dimension of $10h \times 10h$, where $h = 0.5$~mm is the height of the injector lip. A sponge layer of length 5$h$ is put at the end of the domain. A fully structured Cartesian mesh is utilized in this case with 100 grid points across the injector lip. A uniform mesh is used in axial direction. For the region within 3$h$ around the injector lip in transverse direction, a uniform mesh is adopted and stretching is applied with a ratio of 1.02 outside this region. The mesh is stretched in axial direction in the sponge layer. This results in a mesh with a total of $5.2 \times 10^5$ cells.

Adiabatic no-slip wall conditions are applied at the injector lip and adiabatic slip wall conditions are applied for the top and bottom boundaries of the domain. A 1/7th power law for velocity is used for both the LOX and GH2 streams. Pressure outlet boundary conditions are applied after the sponge layer where acoustic waves are suppressed. The LOX stream is injected at a temperature of 100~K, and GH2 is injected at a temperature of 150~K. The pressure is set to 10~MPa which is representative of rocket combustor conditions. Note that the density ratio between LOX and GH2 is about 80. The hybrid scheme using the double-flux model is used with the RS sensor\cite{Hickey2013c, ma2016entropy} set to a value of 0.2. A first-order upwinding scheme is used in the sponge layer. The CFL number is set to 1.0 and no subgrid scale model is used to facilitate comparisons with the reference solutions.

The flamelet table is generated at same conditions, namely $T_\text{ox}$~=~100~K, $T_\text{f}$~=~150~K, and $p$~=~100~bar. Transcritical flamelets are calculated for the entire S-shaped curve at reference conditions. Water is used as the progress variable. Resolutions in mixture fraction and progress variable are both 100 grid points. In mixture fraction dimension, the mesh is uniform from zero to the stoichiometric value using one third of the grid points and then stretches to one with the rest of the grid points. No special treatment is conducted for the resolution in the PBP region for the reason that the current developed FPV model is insensitive to the table resolution at this region as demonstrated in the previous subsection. For the pure mixing case considered in this subsection, the progress variable, $\widetilde{C}$, is set to zero.

\begin{figure}[!t!]
    \centering
        \includegraphics[trim={10 10 10 10},width=9.2cm,clip]{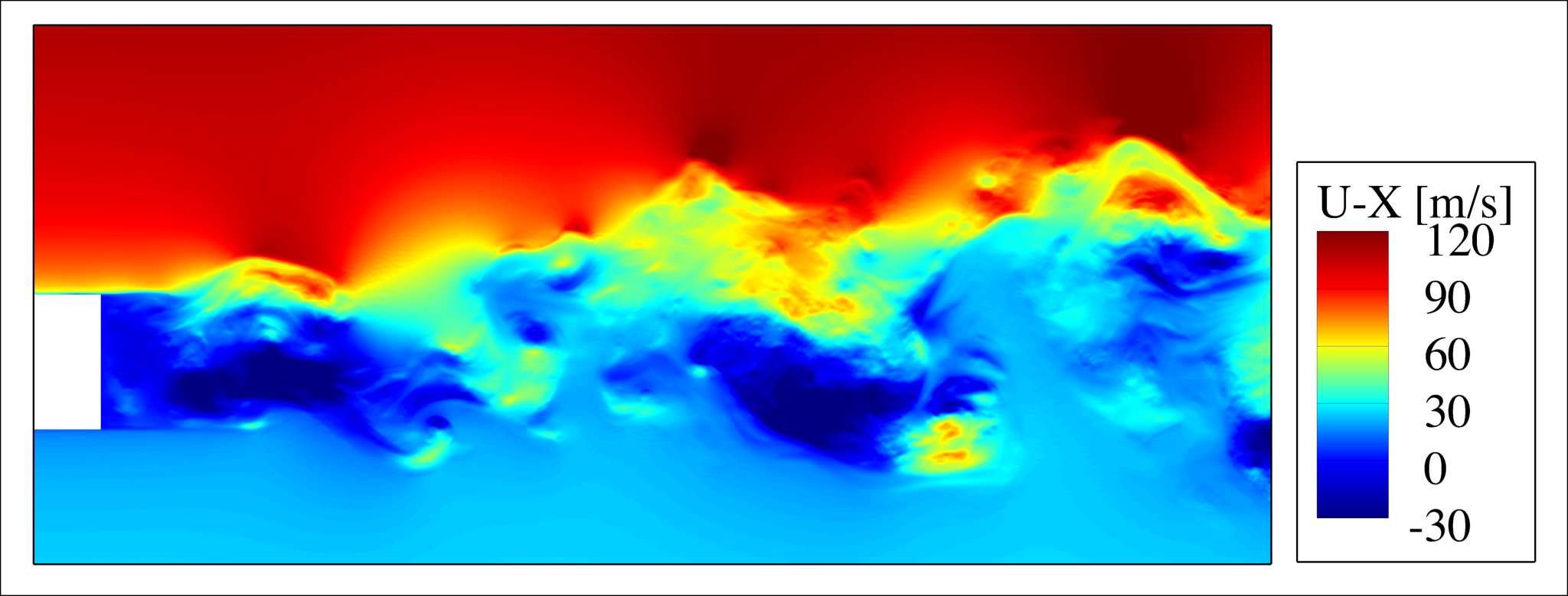}
        \includegraphics[trim={10 10 10 10},width=9.2cm,clip=]{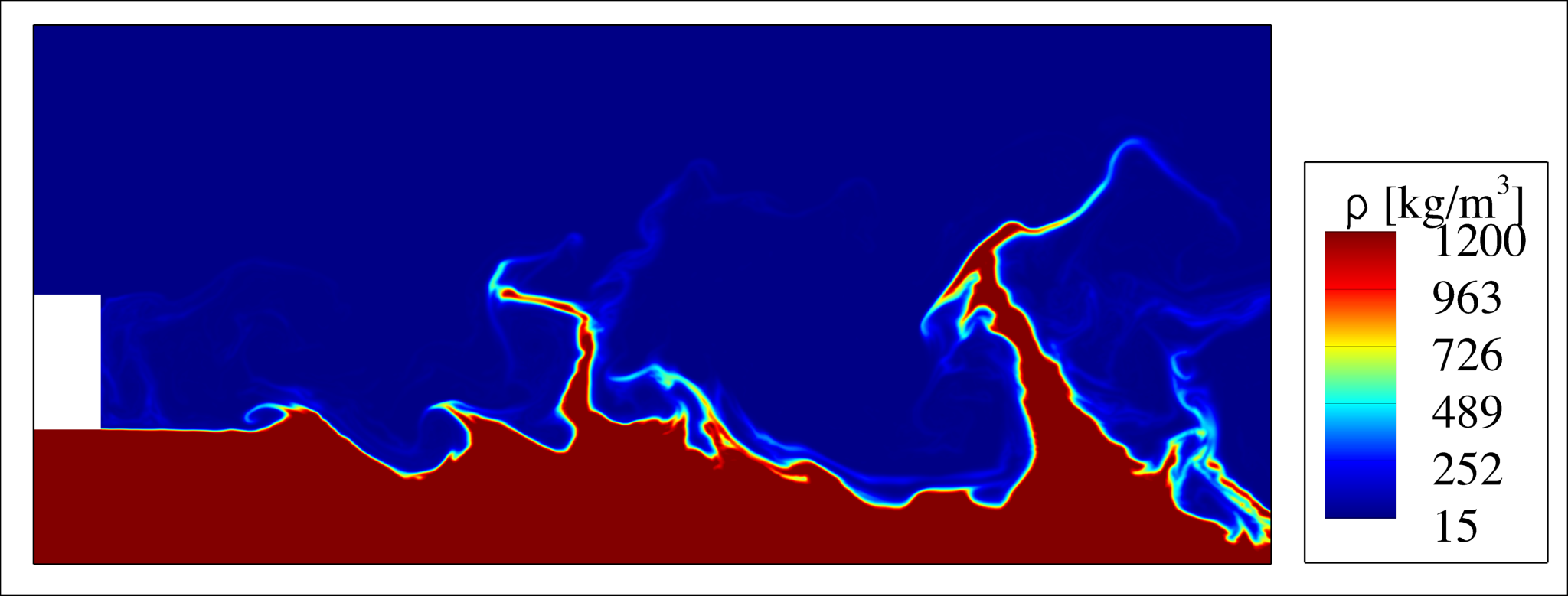}
        \includegraphics[trim={10 10 10 10},width=9.2cm,clip=]{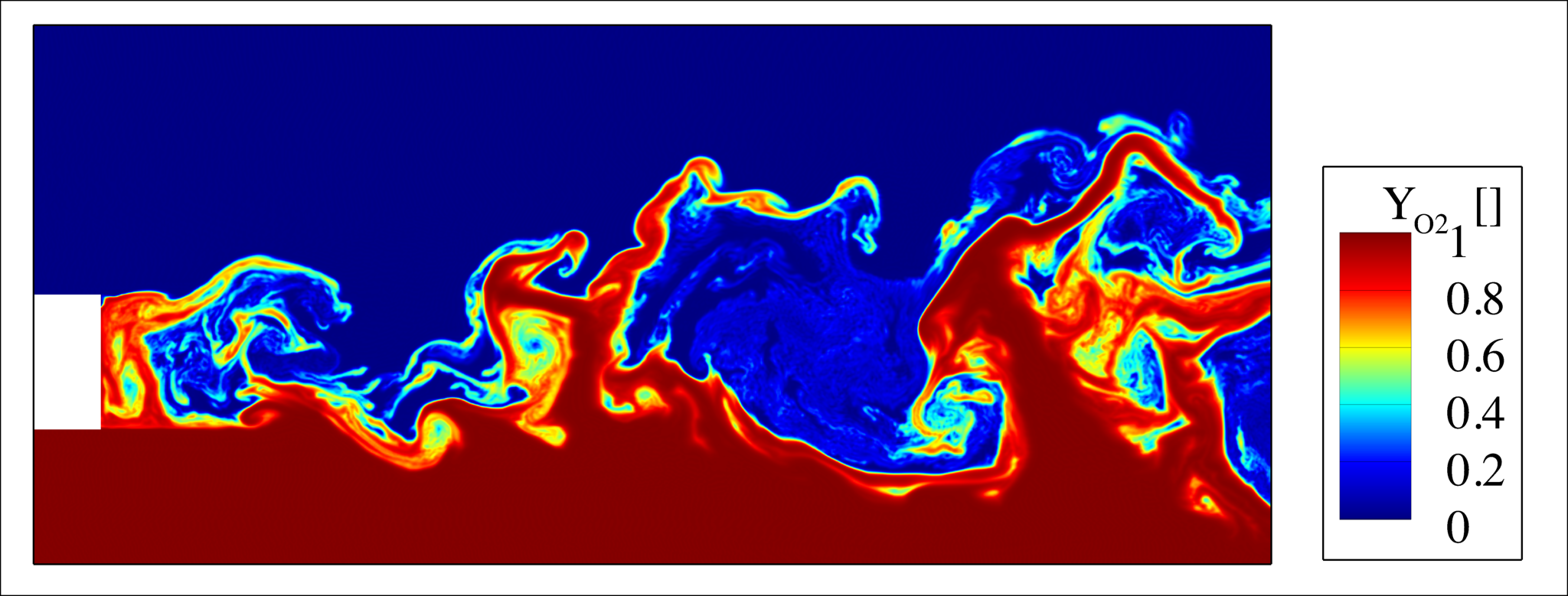}
    \caption{\label{fig:mixing} Instantaneous fields of axial velocity, density, and mass fraction of oxygen from top to bottom for the cryogenic LOX/GH2 mixing case.}
\end{figure}

\Cref{fig:mixing} shows results for instantaneous axial velocity, density, and oxygen mass fraction. Due to the implementation of the proposed FPV model, for pure mixing case, the FPV model is expected to perform almost the same as a multi-component model in which species mass fractions are explicitly solved. Indeed, the mixture fraction for the mixing case considered here acts as the mass fraction of hydrogen. The only difference comes from the evaluation of the thermodynamic quantities which is expected to be small as demonstrated in the previous subsection. The results from the current FPV model is found to give almost the same results both qualitatively and quantitatively as the multi-component model in Ma~et~al.~\cite{ma2016entropy} As can be seen in \cref{fig:mixing}, the flow field is dominated by large vortical structures in the mixing layer and three large vortical structures are separated by waves with a wavelength of approximately $5h$. The predicted structures of vortices are in good agreement with those reported in Ruiz~et~al.~\cite{ruiz2015numerical}. From the density field, ``comb-like" or ``finger-like" structures~\cite{Mayer1998} can clearly be seen, which was also observed through experiments for transcritical mixing under typical rocket engine operating conditions~\cite{Oschwald1999,Chehroudi2002}. 

\begin{figure}[!t!]
    \centering
    \subfigure[Axial velocity]{
        \includegraphics[width=7.2cm,clip=]{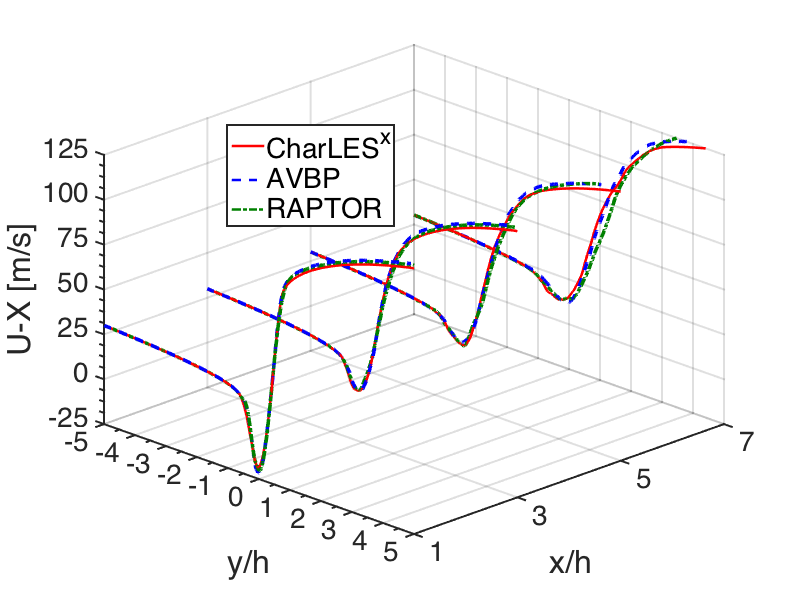}
        \includegraphics[width=7.2cm,clip=]{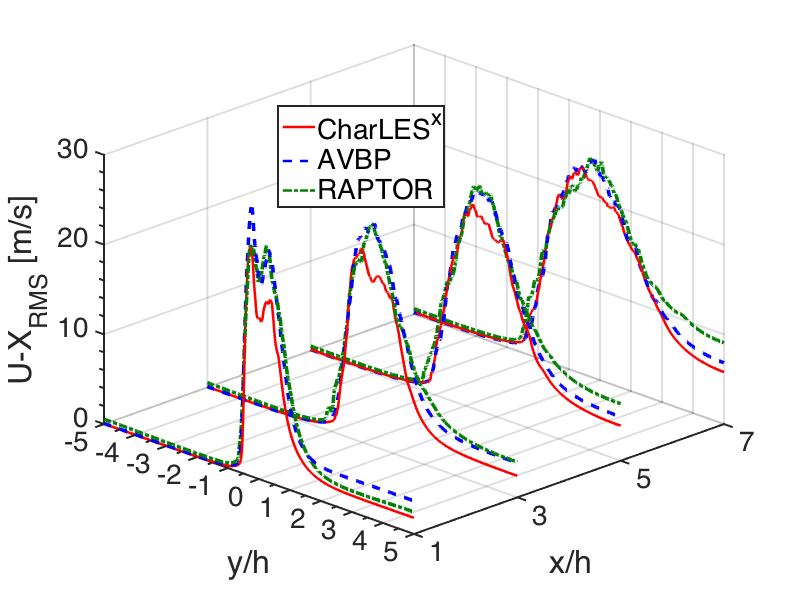}
    }
    \subfigure[Mass fraction of oxygen]{
        \includegraphics[width=7.2cm,clip=]{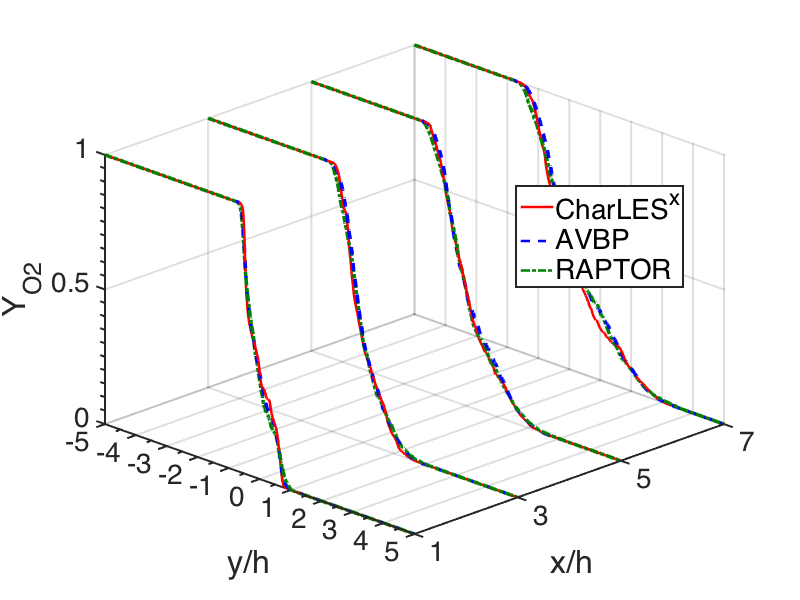}
        \includegraphics[width=7.2cm,clip=]{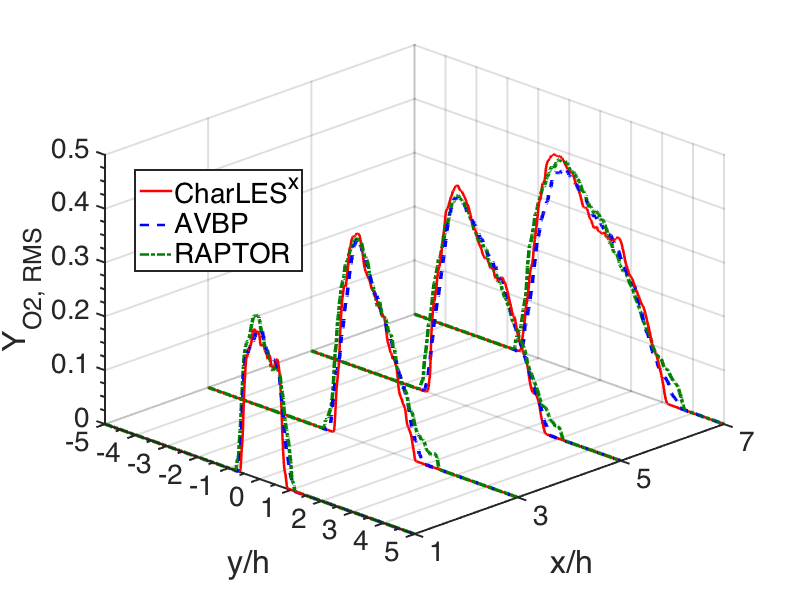}
    }
    \subfigure[Temperature]{
        \includegraphics[width=7.2cm,clip=]{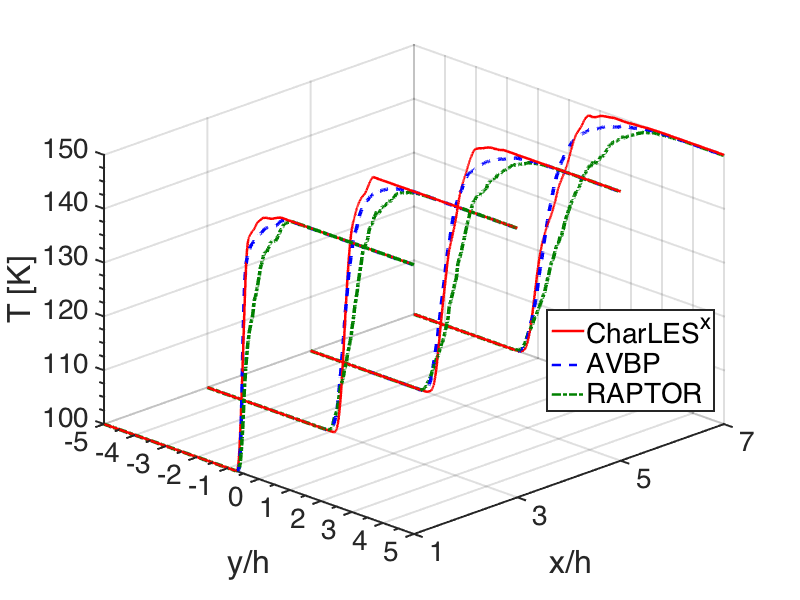}
        \includegraphics[width=7.2cm,clip=]{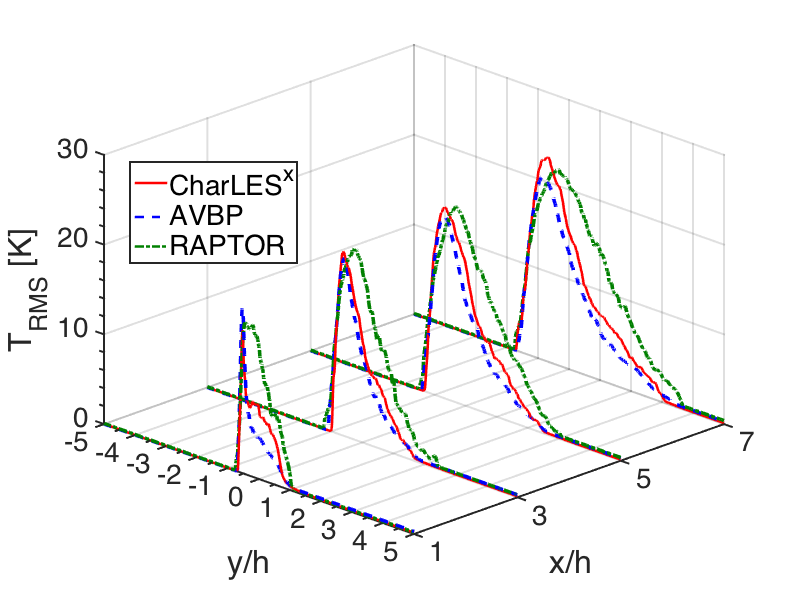}
    }
    \caption{\label{fig:2DMixing} Mean and RMS results for axial velocity, mass fraction of oxygen, and temperature at different transverse cuts in comparison with Ruiz~et~al.~\cite{ruiz2015numerical} for the cryogenic LOX/GH2 mixing case.}
\end{figure}

The simulation results are averaged in time to facilitate quantitative comparisons and 15 flow-through-times are used for the averaging process after steady state is reached. One flow through time corresponds to 0.125~ms~\cite{ruiz2015numerical}. \Cref{fig:2DMixing} shows the results of mean and root mean square (RMS) values for axial velocity, mass fraction of oxygen, and temperature. Statistics at different axial locations ($x/h$~=~1,~3,~5,~7) are plotted as a function of normalized transverse distance. Results from the current solver, {\it CharLES$^{\,x}$}, are compared to those obtained from two other solvers, namely $AVBP$ and $RAPTOR$~\cite{ruiz2015numerical}. The mean axial velocity is in good agreement between the three different solvers, while there are some discrepancies in the RMS values. Results from {\it CharLES$^{\,x}$} show slightly lower RMS values on the GH2 side, especially for the axial location of $x/h = 3$. This is probably due to the different implementation of the sponge layer and outlet boundary conditions adopted by the different solvers. Results for the oxygen mass fraction are almost identical for the three solvers except for the small difference seen at the GH2 side which could be related to the difference seen in the velocity results. Appreciable differences are observed in the temperature statistics. The results from {\it CharLES$^{\,x}$} and {\it AVBP} are similar and show a narrower thermal mixing layer as compared to that of {\it RAPTOR}. Overall, the results obtained from the three solvers are in good agreement, demonstrating the capability of the proposed FPV model and the robustness of the numerical schemes for transcritical mixing cases.

\subsection{Cryogenic LOX/GH2 reacting case}
The cryogenic LOX/GH2 mixing case described in the previous subsection is then ignited to further assess the performance of the proposed FPV model for reacting cases. The computational domain, mesh resolution, boundary conditions, and numerical schemes are kept the same as in the mixing case. The same flamelet table is adopted for the reacting case. As expected, large acoustic waves are generated right after the ignition takes places, and first-order upwinding scheme is used in the sponge layer to ensure that the pressure waves exit the domain and eventually a stable flame is obtained.

\begin{figure}[!t!]
    \centering
    \includegraphics[trim={10 10 10 10},width=9.2cm,clip,frame]{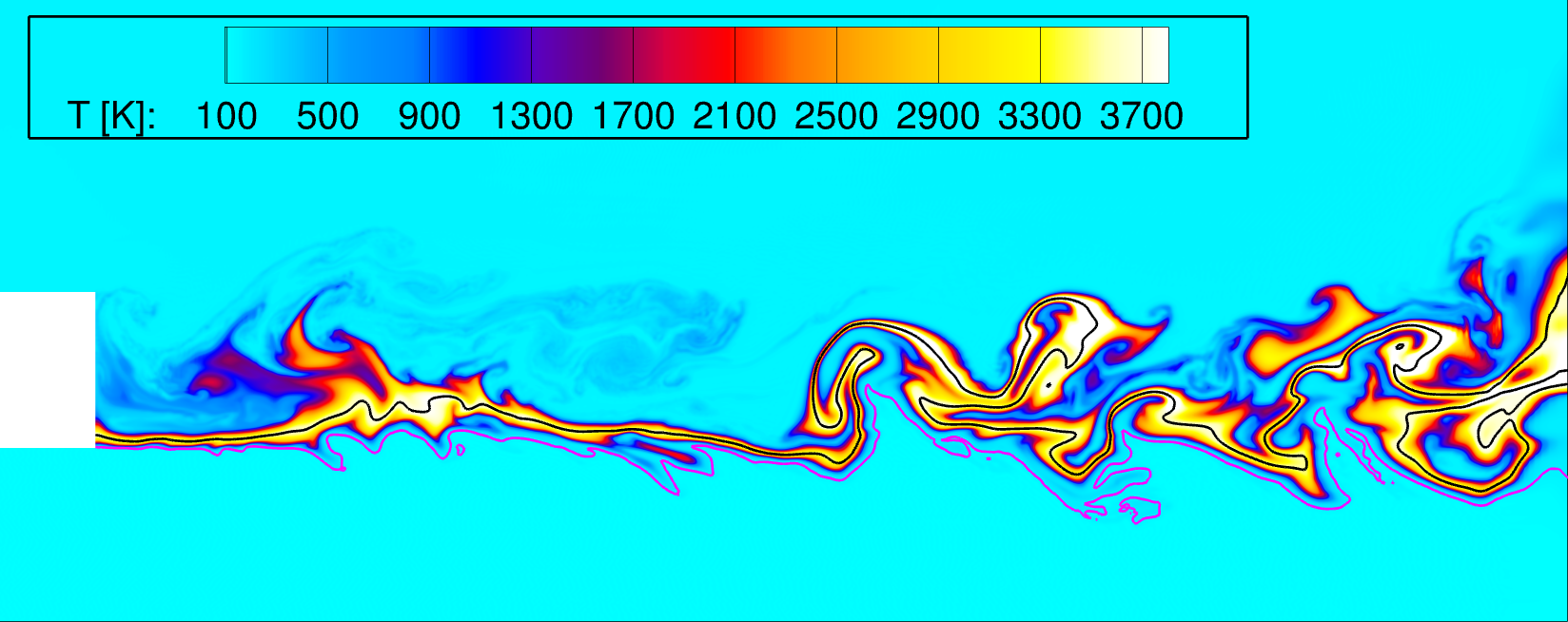} \\[0.5em]
    \includegraphics[trim={10 10 10 10},width=9.2cm,clip,frame]{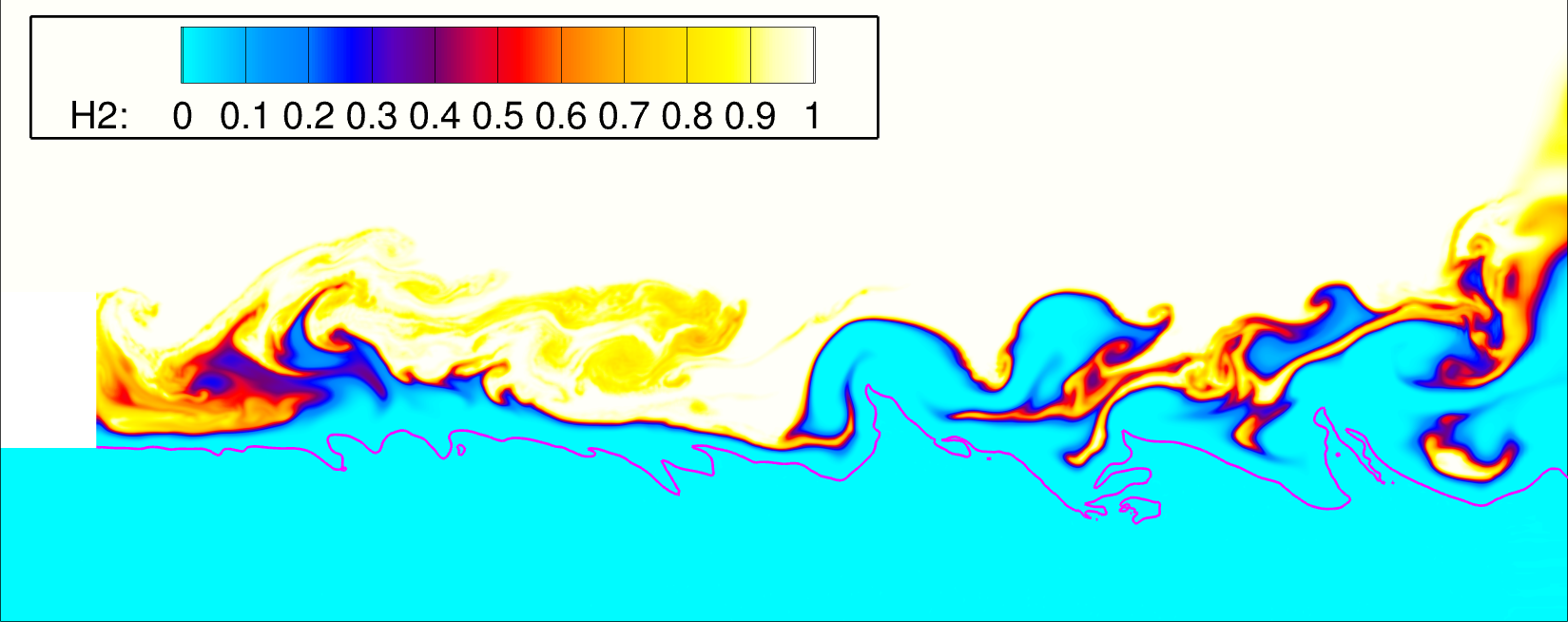} \\[0.5em]
    \includegraphics[trim={10 10 10 10},width=9.2cm,clip,frame]{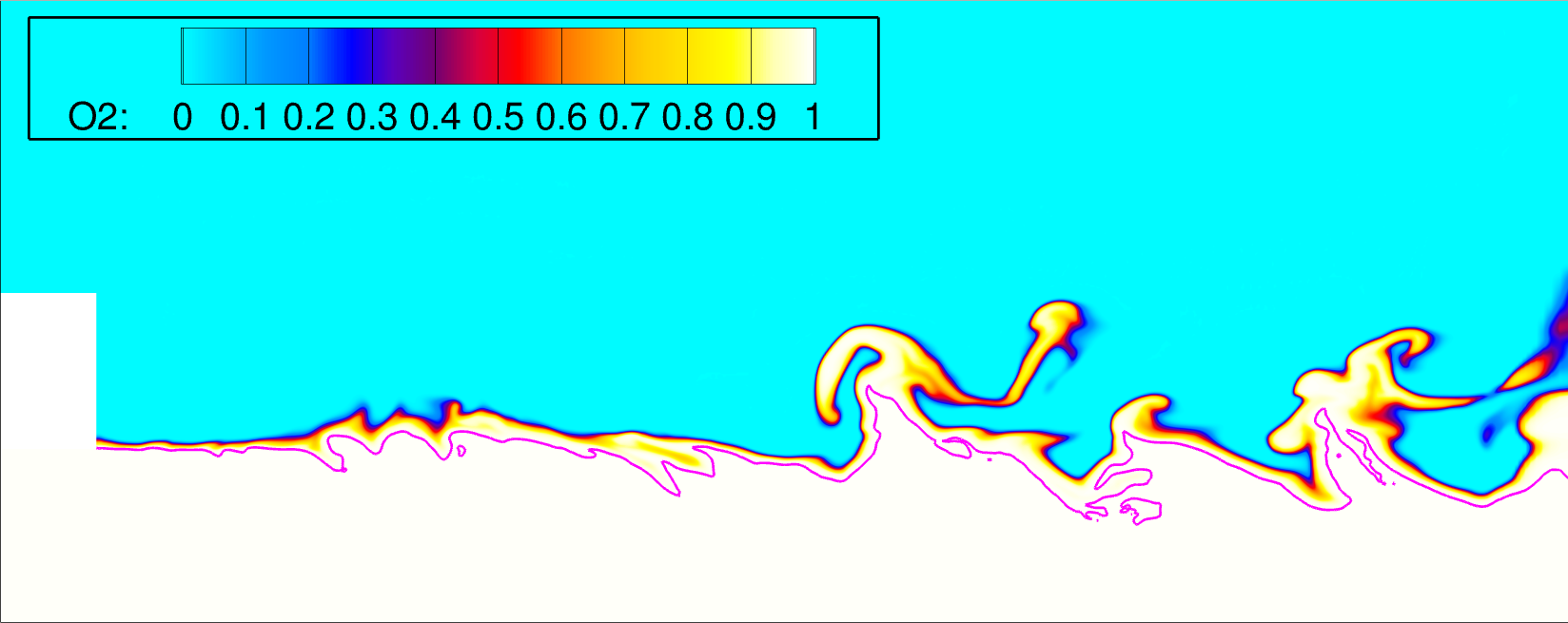} \\[0.5em]
    \includegraphics[trim={10 10 10 10},width=9.2cm,clip,frame]{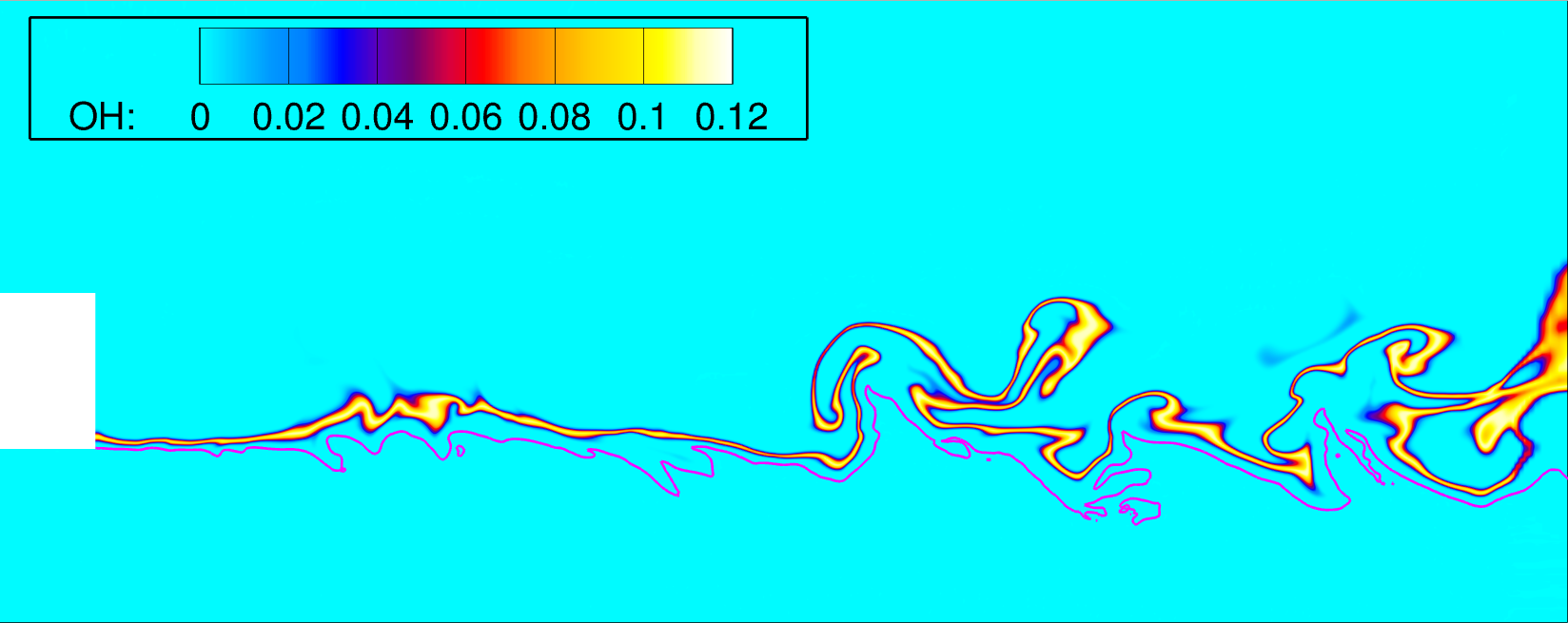}
    \caption{\label{fig:reacting} Instantaneous temperature, mass fractions of H$_2$, O$_2$, and OH from top to bottom for the cryogenic LOX/GH2 reacting case. Black curve corresponds to the stoichiometric mixture fraction, and pink curve indicates the PBP location characterized by the peak of specific heat.}
\end{figure}

\Cref{fig:reacting} shows instantaneous results for temperature, mass fractions of H$_2$, O$_2$, and OH. The black curve in the temperature subfigure in \cref{fig:reacting} indicates the stoichiometric value of mixture fraction. Pink curves in \cref{fig:reacting} correspond to the PBP location which is characterized by the peak of the specific heat capacity. As can be seen from \cref{fig:reacting}, the proposed FPV model is able to predict the transcritical flame robustly due to the double-flux model and entropy flux correction technique adopted in the numerical solver. No spurious oscillations in pressure or velocity were observed. Due to the adiabatic boundary conditions applied at the injector lip, the flame is attached. The flame exhibits laminar behavior close to the injector tip and is more wrinkled and turbulent downstream. The structure of a transcritical flame is found to be similar to that under ideal-gas conditions through flamelet studies. The PBP region is spatially separated from the reaction zone and the real-fluid pseudo-boiling process is found to take place in the region characterized by almost pure oxygen~\cite{banuti2016efficient, banuti2016sub}. This can be seen from the current simulation results that the PBP location as indicated by the pink curve has no interaction with the flame. The transcritical process happens to almost pure oxygen as can be seen from the oxygen mass fraction results in \cref{fig:reacting}.

\begin{figure}[!t!]
    \centering
    \subfigure[Mixing case]{
        \includegraphics[trim={10 10 10 10},width=8.8cm,clip,frame]{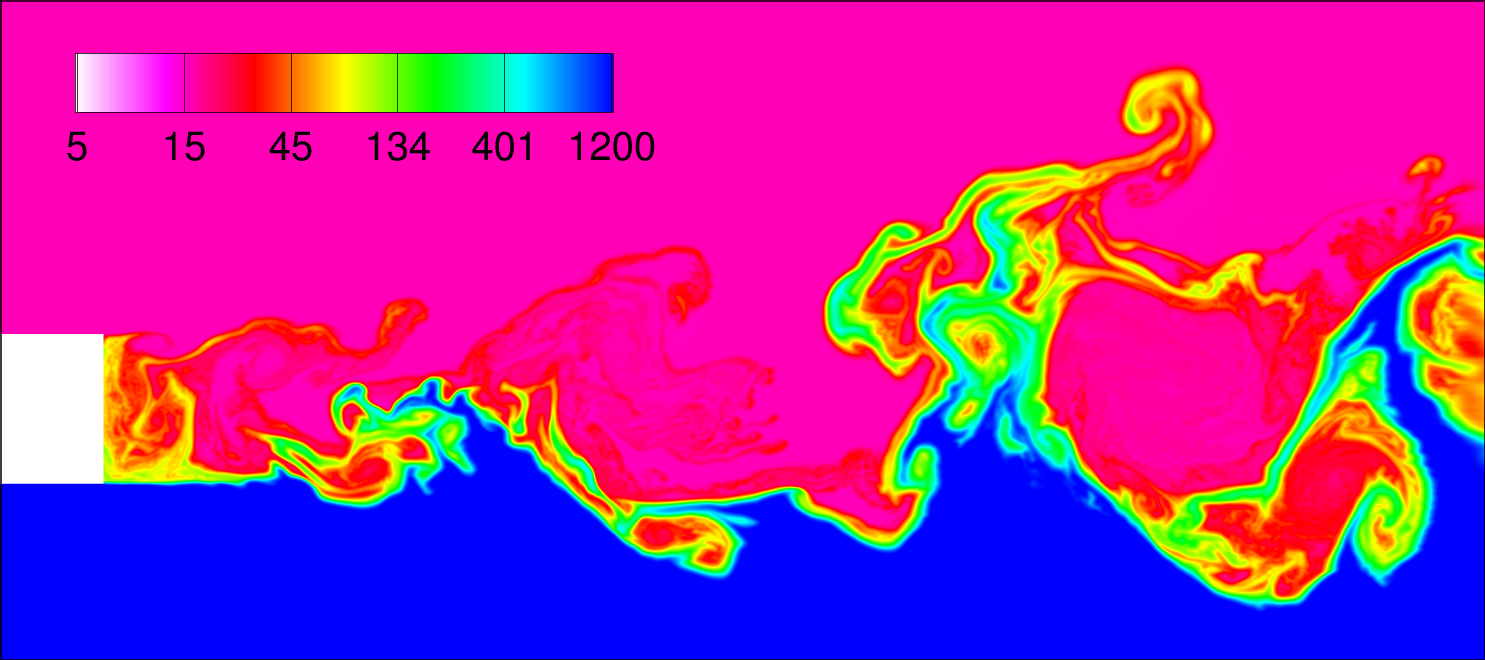}
    }
    \subfigure[Reacting case]{
        \includegraphics[trim={10 10 10 10},width=8.8cm,clip,frame]{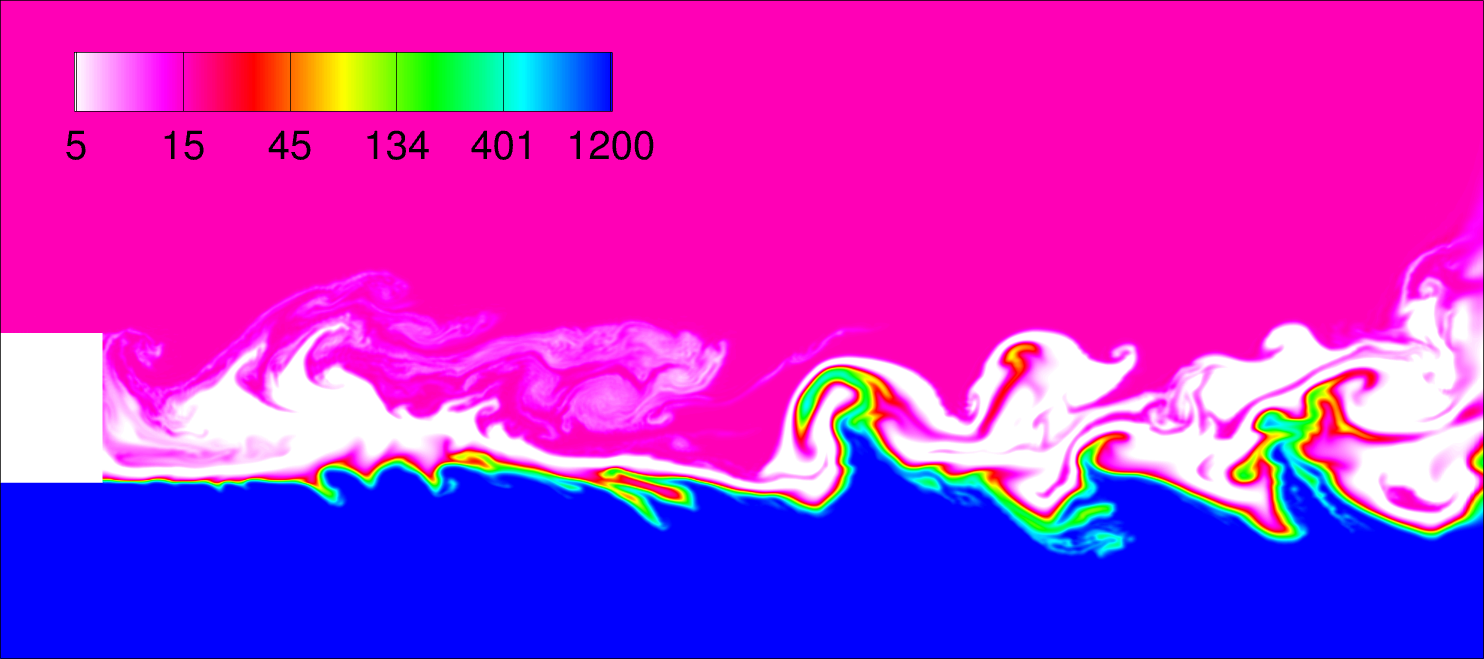}
    }
    \caption{\label{fig:rhocomparison} Instantaneous density fields of the cryogenic LOX/GH2 mixing and reacting cases. Units in kg/m$^3$. Legends in logarithmic scale.}
\end{figure}

\Cref{fig:rhocomparison} compares instantaneous density fields from both the cryogenic mixing and reacting cases. The legends are in logarithmic scale to emphasize the small structures at relatively low density values. As can be clearly seen from \cref{fig:rhocomparison}, the density fields show considerable differences between the two cases. Lower density can be seen in the reacting case due to the flame in between the two streams. The LOX stream in the reacting case shows suppressed vortical structures. Whereas in the mixing case, large vortical structures can be seen and the dense LOX stream penetrates into the GH2 stream by a height of more than $h$ above the injector lip. The heat release from the flame not only separates the two streams which results in an almost pure species pesudo-boiling process, but also suppresses the development of the vortical structures in the mixing layer. This is in consistent with the simulation results by Ruiz~\cite{Ruiz2012}.

The proposed FPV model has been demonstrated to have the capability for transcritical reacting flow simulations. The current numerical scheme is directly applicable to LES of real applications under transcritical conditions.


\section{Conclusions}
A FPV approach is developed for trans- and supercritical combustion simulations in the context of finite volume, fully compressible, explicit solvers. The PR cubic EoS is used for the consistent evaluation of the thermodynamic quantities under transcritical conditions. Capabilities to calculate transcritical flamelet solutions have been implemented in the FlameMaster~\cite{Pitsch2006a} solver and validated against DNS results. The double-flux model developed for transcritical flows~\cite{ma2016entropy} is used to eliminate spurious pressure oscillations caused by the nonlinearity inherent in the real-fluid EoS. A hybrid scheme with entropy-stable flux correction technique~\cite{ma2016entropy} is used to deal with the large density ratio in transcritical combustion cases. For the FPV approach, parameters $a$ and $b$ in the cubic EoS are pre-tabulated for the evaluation of the departure functions and a quadratic model is used to recover the attraction parameter $a$. The ideal-gas values are calculated from a linearized specific heat ratio model.~\cite{saghafian2015efficient} The novelty of the proposed approach lies in the ability to account for pressure and temperature variations from the reference tabulated values using a computationally tractable pre-tabulated combustion chemistry in a thermodynamically consistent fashion. An {\it a priori} analysis was conducted to assess the performance of the proposed transcritical FPV approach and it is shown that the assumption that the composition is a weak function of the reference conditions is valid under transcritical conditions, and the currently developed FPV approach can accurately recover the real-fluid and ideal-gas thermodynamic quantities despite perturbations in temperature and pressure. The solution of the laminar flamelets in mixture fraction space and the chemistry tabulation requires special considerations in order to account for the full non-linear effects in transcritical flows. The transcritical FPV approach works robustly and accurately even with an insufficient resolution in mixture fraction in the table. Cryogenic LOX/GH2 mixing and reacting cases are performed to demonstrate the capability of the proposed approach in multidimensional simulations. The current combustion model and numerical schemes are directly applicable to LES of real applications under transcritical conditions.

\section*{Acknowledgments}
Financial support through NASA with award numbers NNX14CM43P and NNM13AA11G are gratefully acknowledged. The authors would like to thank Dr. Guilhem Lacaze for sharing data for comparison.


\begin{thebibliography}{10}
\newcommand{\enquote}[1]{``#1''}

\bibitem{Chehroudi2012}
Chehroudi, B., \enquote{Recent experimental efforts on high-pressure
  supercritical injection for liquid rockets and their implications,} {\em Int.
  J. Aero. Eng.\/}, Vol.~121802, 2012, pp.~31.

\bibitem{Mayer1998}
Mayer, W. O.~H., Schik, A. H.~A., Vielle, B., Chauveau, C., Goekalp, I.,
  Talley, D.~G., and Woodward, R.~D., \enquote{Atomization and breakup of
  cryogenic propellants under high-pressure subcritical and supercritical
  conditions,} {\em J. Prop. Power\/}, Vol.~14, No.~5, 1998, pp.~835--842.

\bibitem{Oschwald1999}
Oschwald, M. and Schik, A., \enquote{Supercritical nitrogen free jet
  investigated by spontaneous {R}aman scattering,} {\em Exp. Fluids\/},
  Vol.~27, No.~6, 1999, pp.~497--506.

\bibitem{Chehroudi2002}
Chehroudi, B., Talley, D., and Coy, E., \enquote{Visual characteristics and
  initial growth rates of round cryogenic jets at subcritical and supercritical
  pressures,} {\em Phys. Fluids\/}, Vol.~14, 2002, pp.~850.

\bibitem{Habiballah2006}
Habiballah, M., Orain, M., Grisch, F., Vingert, L., and Gicquel, P.,
  \enquote{Experimental studies of high-pressure cryogenic flames on the
  Mascotte facility,} {\em Combust. Sci. Technol.\/}, Vol.~178, 2006,
  pp.~101--128.

\bibitem{Davis2007}
Davis, D.~W. and Chehroudi, B., \enquote{Measurements in an acoustically driven
  coaxial jet under sub-, near-, and supercritical conditions,} {\em J. Propul.
  Power\/}, Vol.~23, No.~2, 2007, pp.~364--374.

\bibitem{Dahms2013}
Dahms, R.~N. and Oefelein, J.~C., \enquote{On the transition between two-phase
  and single-phase interface dynamics in multicomponent fluids at supercritical
  pressures,} {\em Phys. Fluids\/}, Vol.~25, 2013, pp.~092103.

\bibitem{Ivancic2002}
Ivancic, B. and Mayer, W., \enquote{Time- and length scales of combustion in
  liquid rocket thrust chambers,} {\em J. Propul. Power\/}, Vol.~18, No.~2,
  2002, pp.~247--253.

\bibitem{Ribert2008}
Ribert, G., Zong, N., Yang, V., Pons, L., Darabiha, N., and Candel, S.,
  \enquote{Counterflow diffusion flames of general fluids: Oxygen/hydrogen
  mixtures,} {\em Combust. Flame\/}, Vol.~154, 2008, pp.~319--330.

\bibitem{Lacaze2012}
Lacaze, G. and Oefelein, J.~C., \enquote{A non-premixed combustion model based
  on flame structure analysis at supercritical pressures,} {\em Combust.
  Flame\/}, Vol.~158, 2012, pp.~2087--2103.

\bibitem{Pitsch2006}
Pitsch, H., \enquote{Large-Eddy Simulation of Turbulent Combustion,} {\em Annu.
  Rev. Fluid Mech\/}, Vol.~38, 2006.

\bibitem{Zong2008}
Zong, N., Ribert, G., and Yang, V., \enquote{A flamelet approach for modeling
  of liquid oxygen (LOX)/methane flames at supercritical pressures,} {\em AIAA
  paper 2008-946\/}, 2008.

\bibitem{Cutrone2010}
Cutrone, L., Palma, P.~D., Pascazio, G., and Napolitano, M., \enquote{A {RANS}
  flamelet/progress-variable method for computing reacting flows of real-gas
  mixtures,} {\em Comput. Fluids\/}, Vol.~39, No.~3, 2010, pp.~485 -- 498.

\bibitem{Kim2011a}
Kim, T., Kim, Y., and Kim, S.-K., \enquote{Numerical analysis of gaseous
  hydrogen/liquid oxygen flamelet at supercritical pressures,} {\em Int. J.
  Hydrogen Energy\/}, Vol.~36, No.~10, 2011, pp.~6303 -- 6316.

\bibitem{huo2014general}
Huo, H., Wang, X., and Yang, V., \enquote{A general study of counterflow
  diffusion flames at subcritical and supercritical conditions: oxygen/hydrogen
  mixtures,} {\em Combust. Flame\/}, Vol.~161, No.~12, 2014, pp.~3040--3050.

\bibitem{banuti2016sub}
Banuti, D.~T., Ma, P.~C., Hickey, J.-P., and Ihme, M., \enquote{Sub- or
  supercritical? A flamelet analysis of high pressure rocket propellant
  injection,} {\em AIAA Paper 2016-4789\/}, 2016.

\bibitem{Pitsch1998}
Pitsch, H., Chen, M., and Peters, N., \enquote{Unsteady flamelet modeling of
  turbulent hydrogen-air diffusion flames,} {\em Twenty-Seventh Symposium on
  Combustion, The Combustion Institute\/}, 1998.

\bibitem{wu2015pareto}
Wu, H., See, Y.~C., Wang, Q., and Ihme, M., \enquote{A {Pareto}-efficient
  combustion framework with submodel assignment for predicting complex flame
  configurations,} {\em Combust. Flame\/}, Vol.~162, No.~11, 2015,
  pp.~4208--4230.

\bibitem{wu2016compliance}
Wu, H. and Ihme, M., \enquote{Compliance of combustion models for turbulent
  reacting flow simulations,} {\em Fuel\/}, Vol.~186, 2016, pp.~853--863.

\bibitem{Candel2006}
Candel, S., Juniper, M., Singla, G., Scouflaire, P., and Rolon, C.,
  \enquote{Structure and dynamics of cryogenic flames at supercritical
  pressure,} {\em Combust. Sci. Technol.\/}, Vol.~178, No. 1-3, 2006,
  pp.~161--192.

\bibitem{Yang2007}
Yang, B., Cuoco, F., and Oschwald, M., \enquote{Atomization and flames in
  {LOX}/{H2}- and {LOX}/{CH4}- spray combustion,} {\em J. Propul. Power\/},
  Vol.~23, No.~4, 2007, pp.~763--771.

\bibitem{Pierce2004}
Pierce, C. and Moin, P., \enquote{Progress-variable approach for large-eddy
  simulation of non-premixed turbulent combustion,} {\em J. Fluid Mech.\/},
  Vol.~504, 2004, pp.~73--97.

\bibitem{Ihme2005}
Ihme, M., Cha, C., and Pitsch, H., \enquote{Prediction of local extinction and
  re-ignition effects in non-premixed turbulent combustion using a
  flamelet/progress variable approach,} {\em P. Comb. Inst.\/}, Vol.~30, 2005,
  pp.~793--800.

\bibitem{Giorgi2014}
Giorgi, M. G.~D., Sciolti, A., and Ficarella, A., \enquote{Application and
  comparison of different combustion models of high pressure LOX/CH4 jet
  flames,} {\em Energies\/}, Vol.~7, 2014, pp.~477--497.

\bibitem{Oefelein1998}
Oefelein, J.~C. and Yang, V., \enquote{Modeling high-pressure mixing and
  combustion processes in liquid rocket engines,} {\em J. Prop. Power\/},
  Vol.~14, No.~5, 1998, pp.~843--857.

\bibitem{Schmitt2011}
Schmitt, T., Mery, Y., Boileau, M., and Candel, S., \enquote{Large-eddy
  simulation of oxygen/methane flames under transcritical conditions,} {\em P.
  Comb. Inst.\/}, Vol.~33, No.~1, 2011, pp.~1383--1390.

\bibitem{masquelet2010large}
Masquelet, M. and Menon, S., \enquote{Large-eddy simulation of flame-turbulence
  interactions in a shear coaxial injector,} {\em J. Propul. Power\/}, Vol.~26,
  No.~5, 2010, pp.~924--935.

\bibitem{Selle2007}
Selle, A., Okong'o, N.~A., Bellan, J., and Harstad, K.~G., \enquote{Modelling
  of subgrid-scale phenomena in supercritical transitional mixing layers: an a
  priori study,} {\em J. Fluid Mech.\/}, Vol.~593, No. 57--91, 2007.

\bibitem{Huo2013b}
Huo, H. and Yang, V., \enquote{Sub-grid scale models for large-eddy simulation
  of supercritical combustion,} {\em AIAA paper 2013-0706\/}, 2013.

\bibitem{Petit2013b}
Petit, X., Ribert, G., Lartigue, G., and Domingo, P., \enquote{Large-eddy
  simulation of supercritical fluid injection,} {\em The Journal of
  Supercritical Fluids\/}, Vol.~84, 2013, pp.~61--73.

\bibitem{Terashima2011}
Terashima, H., Kawai, S., and Yamanishi, N., \enquote{High-resolution numerical
  method for supercritical flows with large density variations,} {\em AIAA
  J.\/}, Vol.~49, No.~12, 2011, pp.~2658--2672.

\bibitem{Terashima2012a}
Terashima, H. and Koshi, M., \enquote{Approach for simulating gas/liquid-like
  flows under supercritical pressures using a high-order central differencing
  scheme,} {\em J. Comp. Phys.\/}, Vol.~231, No.~20, 2012, pp.~6907 -- 6923.

\bibitem{Terashima2013}
Terashima, H. and Koshi, M., \enquote{Strategy for simulating supercritical
  cryogenic jets using high-order schemes,} {\em Comput. Fluids\/}, Vol.~85,
  2013.

\bibitem{Lacaze2013}
Lacaze, G. and Oefelein, J., \enquote{Modeling of high density gradient flows
  at supercritical pressures,} {\em AIAA paper 2013-3717\/}, 2013.

\bibitem{ma2014supercritical}
Ma, P.~C., Bravo, L., and Ihme, M., \enquote{Supercritical and transcritical
  real-fluid mixing in diesel engine applications,} {\em Proceedings of the
  Summer Program, Center for Turbulence Research, Stanford University\/}, 2014,
  pp.~99--108.

\bibitem{ma2016numerical}
Ma, P.~C., Lv, Y., and Ihme, M., \enquote{Numerical methods to prevent pressure
  oscillations in transcritical flows,} {\em Annual Research Brief, Center for
  Turbulence Research, Stanford University\/}, 2016, pp.~223--234.

\bibitem{ma2016entropy}
Ma, P.~C., Lv, Y., and Ihme, M., \enquote{An entropy-stable hybrid scheme for
  simulations of transcritical real-fluid flows,} {\em J. Comput. Phys.\/},
  Vol.~340, 2017, pp.~330--357.

\bibitem{Okongo2002}
Okong'o, N. and Bellan, J., \enquote{Consistent boundary conditions for
  multicomponent real gas mixtures based on characteristic waves,} {\em J.
  Comp. Phys.\/}, Vol.~176, No.~2, 2002, pp.~330--344.

\bibitem{Coussement2013}
Coussement, A., Gicquel, O., Fiorina, B., Degrez, G., and Darabiha, N.,
  \enquote{Multicomponent real gas 3-D-NSCBC for direct numerical simulation of
  reactive compressible viscous Flows,} {\em J. Comp. Phys.\/}, Vol.~245, 2013,
  pp.~259--280.

\bibitem{petit2015framework}
Petit, X., Ribert, G., and Domingo, P., \enquote{Framework for real-gas
  compressible reacting flows with tabulated thermochemistry,} {\em J.
  Supercrit. Fluids\/}, Vol.~101, 2015, pp.~1--16.

\bibitem{Miller2001}
Miller, R., Harstad, K., and Bellan, J., \enquote{Direct numerical simulations
  of supercritical fluid mixing layers applied to heptane-nitrogen,} {\em J.
  Fluid Mech.\/}, Vol.~436, No.~6, 2001, pp.~1--39.

\bibitem{Okongo2002a}
Okong'o, N.~A. and Bellan, J., \enquote{Direct numerical simulation of a
  transitional supercritical binary mixing layer: heptane and nitrogen,} {\em
  J. Fluid Mech.\/}, Vol.~464, 2002, pp.~1--34.

\bibitem{Okongo2003a}
Okong'o, N. and Bellan, J., \enquote{Real-gas effects on mean flow and temporal
  stability of binary-species mixing layers,} {\em AIAA J.\/}, Vol.~41, No.~12,
  2003, pp.~2429--2443.

\bibitem{Taskinoglu2010}
Taskinoglu, E. and Bellan, J., \enquote{A posteriori study using a DNS database
  describing fluid disintegration and binary-species mixing under supercritical
  pressure: heptane and nitrogen,} {\em J. Fluid Mech.\/}, Vol.~645, 2010,
  pp.~211--254.

\bibitem{miller2001direct}
Miller, R.~S., Harstad, K.~G., and Bellan, J., \enquote{Direct numerical
  simulations of supercritical fluid mixing layers applied to
  heptane--nitrogen,} {\em J. Fluid Mech.\/}, Vol.~436, 2001, pp.~1--39.

\bibitem{peng1976new}
Peng, D.-Y. and Robinson, D.~B., \enquote{A new two-constant equation of
  state,} {\em Ind. Eng. Chem. Res.\/}, Vol.~15, No.~1, 1976, pp.~59--64.

\bibitem{poling2001properties}
Poling, B.~E., Prausnitz, J.~M., and O'Connell, J.~P., {\em The properties of
  gases and liquids\/}, McGraw-Hill New York, 2001.

\bibitem{ely1981prediction}
Ely, J.~F. and Hanley, H., \enquote{Prediction of transport properties. 1.
  {Viscosity} of fluids and mixtures,} {\em Ind. Eng. Chem. Res.\/}, Vol.~20,
  No.~4, 1981, pp.~323--332.

\bibitem{ely1983prediction}
Ely, J.~F. and Hanley, H., \enquote{Prediction of transport properties. 2.
  {Thermal} conductivity of pure fluids and mixtures,} {\em Ind. Eng. Chem.
  Res.\/}, Vol.~22, No.~1, 1983, pp.~90--97.

\bibitem{harstad1997efficient}
Harstad, K.~G., Miller, R.~S., and Bellan, J., \enquote{Efficient high-pressure
  state equations,} {\em AIChE J.\/}, Vol.~43, No.~6, 1997, pp.~1605--1610.

\bibitem{meng2003unified}
Meng, H. and Yang, V., \enquote{A unified treatment of general fluid
  thermodynamics and its application to a preconditioning scheme,} {\em J.
  Comput. Phys.\/}, Vol.~189, No.~1, 2003, pp.~277--304.

\bibitem{chung1984}
Chung, T.~H., Lee, L.~L., and Starling, K.~E., \enquote{Applications of kinetic
  gas theories and multiparameter correlation for prediction of dilute gas
  viscosity and thermal conductivity,} {\em Ind. Eng. Chem. Res.\/}, Vol.~23,
  No.~1, 1984, pp.~8--13.

\bibitem{chung1988}
Chung, T.~H., Ajlan, M., Lee, L.~L., and Starling, K.~E., \enquote{Generalized
  multiparameter correlation for nonpolar and polar fluid transport
  properties,} {\em Ind. Eng. Chem. Res.\/}, Vol.~27, No.~4, 1988,
  pp.~671--679.

\bibitem{Hickey2013c}
Hickey, J.-P., Ma, P.~C., Ihme, M., and Thakur, S., \enquote{Large eddy
  simulation of shear coaxial rocket injector: {R}eal fluid effects,} {\em AIAA
  paper 2013-4071\/}, 2013.

\bibitem{ruiz2015numerical}
Ruiz, A., Lacaze, G., Oefelein, J., Mari, R., Cuenot, B., Selle, L., and
  Poinsot, T., \enquote{Numerical benchmark for high-{Reynolds}-number
  supercritical flows with large density gradients,} {\em AIAA J.\/}, Vol.~54,
  No.~5, 2015, pp.~1--16.

\bibitem{Peters1983}
Peters, N., \enquote{Local quenching due to flame stretch and non-premixed
  turbulent combustion,} {\em Combust. Sci. Technol.\/}, Vol.~30, No. 1-6,
  1983, pp.~1--17.

\bibitem{Pitsch2006a}
Pitsch, H., \enquote{Creating a flamelet library for the steady flamelet model
  of the flamelet/progress variable approach,} Tech. rep., Stanford University,
  2006.

\bibitem{Burke2012}
Burke, M.~P., Chaos, M., Ju, Y., Dryer, F.~L., and Klippenstein, S.~J.,
  \enquote{Comprehensive {H}2/{O}2 kinetic model for high-pressure combustion,}
  {\em Int. J. Chem. Kinet.\/}, Vol.~44, No.~7, 2012, pp.~444--474.

\bibitem{banuti2015crossing}
Banuti, D.~T., \enquote{Crossing the Widom-line--Supercritical pseudo-boiling,}
  {\em J. Supercrit. Fluids\/}, Vol.~98, 2015, pp.~12--16.

\bibitem{saghafian2015efficient}
Saghafian, A., Terrapon, V.~E., and Pitsch, H., \enquote{An efficient
  flamelet-based combustion model for compressible flows,} {\em Combust.
  Flame\/}, Vol.~162, No.~3, 2015, pp.~652--667.

\bibitem{soave1972equilibrium}
Soave, G., \enquote{Equilibrium constants from a modified {Redlich-Kwong}
  equation of state,} {\em Chem. Eng. Sci.\/}, Vol.~27, No.~6, 1972,
  pp.~1197--1203.

\bibitem{gottlieb2001strong}
Gottlieb, S., Shu, C.-W., and Tadmor, E., \enquote{Strong stability-preserving
  high-order time discretization methods,} {\em SIAM Rev.\/}, Vol.~43, No.~1,
  2001, pp.~89--112.

\bibitem{terashima2012approach}
Terashima, H. and Koshi, M., \enquote{Approach for simulating gas--liquid-like
  flows under supercritical pressures using a high-order central differencing
  scheme,} {\em J. Comput. Phys.\/}, Vol.~231, No.~20, 2012, pp.~6907--6923.

\bibitem{strang1968construction}
Strang, G., \enquote{On the construction and comparison of difference schemes,}
  {\em SIAM J. Num. Anal.\/}, Vol.~5, No.~3, 1968, pp.~506--517.

\bibitem{pecnik2012reynolds}
Pecnik, R., Terrapon, V.~E., Ham, F., Iaccarino, G., and Pitsch, H.,
  \enquote{Reynolds-averaged Navier-Stokes simulations of the HyShot II
  scramjet,} {\em AIAA J.\/}, Vol.~50, No.~8, 2012, pp.~1717--1732.

\bibitem{banuti2016efficient}
Banuti, D.~T., Hannemann, V., Hannemann, K., and Weigand, B., \enquote{An
  efficient multi-fluid-mixing model for real gas reacting flows in liquid
  propellant rocket engines,} {\em Combust. Flame\/}, Vol.~168, 2016,
  pp.~98--112.

\bibitem{Ruiz2012}
Ruiz, A., {\em Unsteady numerical simulations of transcritical turbulent
  combustion in liquid rocket engines\/}, Ph.D. thesis, Institut National
  Polytechnique de Toulouse, France, 2012.

\end{thebibliography}

\end{document}